\documentclass[final]{ar2e}
\usepackage{ulem}  
\begin{document}
\input epsf.tex    

\input psfig.sty
\jname{Annu. Rev. Nucl. Part. Sci.}
\jyear{2004}
\jvol{54:217}
\title{ELECTROMAGNETIC FORM FACTORS OF THE NUCLEON
AND COMPTON SCATTERING}

\markboth{C Hyde-Wright \& K de Jager}
{Nucleon Form Factors and Compton Scattering}

\author{{Charles Earl Hyde-Wright}
\affiliation{Department of Physics, Old Dominion University, Norfolk, Virginia 23529, email: chyde@odu.edu}
Kees de Jager
\affiliation{Thomas Jefferson National Accelerator Facility, Newport News, Virginia 23606, email: kees@jlab.org}}


\begin{abstract}
We review the experimental and theoretical status 
of elastic electron scattering and elastic low-energy photon
scattering (with both real and virtual
photons) from the nucleon.  As a consequence of new experimental
facilities and new theoretical insights, these subjects are
advancing with unprecedented precision.  These
reactions provide many important insights into the spatial distributions
and correlations of quarks in the nucleon.
\end{abstract}
	
\maketitle

\def\Journal#1#2#3#4{{#1} {#2}:#3 (#4)}
\def\NCA{\it Nuovo Cimento}
\def\NIM{\it Nucl. Instrum. Methods}
\def\NIMA{{\it Nucl. Instrum. Methods} A}
\def\EPJ{{\it Eur. Phys. Jour.} A}
\def\JPG{\it J. Phys. G: Nucl. Part. Phys.}
\def\NPA{{\it Nucl. Phys.} A}
\def\NPB{{\it Nucl. Phys.} B}
\def\PLB{{\it Phys. Lett.}  B}
\def\PRL{\it Phys. Rev. Lett.}
\def\PRC{{\it Phys. Rev.}  C}
\def\PRD{{\it Phys. Rev.} D}
\def\RMP{\it Rev. Mod. Phys.}
\def\ZPA{{\it Z. Phys.} A}
\def\ZPC{{\it Z. Phys.} C}

\section{General Introduction}

Although nucleons account for nearly all the visible mass in the universe, they have a complicated structure that is still incompletely understood. The first indication that nucleons have an internal structure, was the 1933 measurement of the proton magnetic moment by Frisch \& 
Stern\cite{stern}. The investigation of the spatial structure of the nucleon was initiated by the HEPL (Stanford) experiments in the 1950s, for which Hofstadter was awarded the 1961 Nobel prize. Several volumes of the {\it Annual Review of Nuclear Science}\cite{hofstadter,wilson} reviewed the status of this field. The recent revival of its experimental study through the operational implementation of novel instrumentation has instigated a strong theoretical interest.

Nucleon electro-magnetic form factors (EMFFs) are optimally studied through the exchange of a virtual photon, in elastic electron-nucleon scattering. The momentum transfer to the nucleon can
be selected to probe different scales of the nucleon, from integral properties such as the 
charge radius to scaling properties of its internal constituents. Polarization instrumentation,
polarized beams and targets, and the measurement of the recoil polarization have been
essential in the accurate separation of the charge and magnetic form factors and in studies
of the elusive neutron charge form factor.

Exclusive Compton scattering on a nucleon refers to the reactions
$\gamma N \rightarrow \gamma' N'$, where either photon may be real or virtual.
In general, the Compton amplitude depends on the full complexity
of the dynamics of the excitation spectrum of the nucleon.  However,
in a number of special kinematic domains, observables with a particularly
simple interpretation can be extracted
from the Compton amplitude.  
In this review, we present the experimental and theoretical
status of   real (RCS) and virtual (VCS) Compton scattering
for the study of generalized polarizabilities,
which measure the spatial distribution
of the response of the nucleon
to external electromagnetic fields.  A thorough discussion of the
rapid developments in  high energy Compton scattering,
in both the deep virtual and hard scattering limits,
is beyond the scope of this review.

\newcommand{\GeV}{GeV$^2$}
\newcommand{\deut}{$^2$H}
\newcommand{\pdeut}{$\stackrel{\rightarrow}{^2\rm{H}}$}
\newcommand{\he}{$^3\rm{He}$}
\newcommand{\phe}{$\stackrel{\rightarrow}{^3\rm{He}}$}
\newcommand{\pDeen}{\pdeut($\vec{e},e^\prime n$)}
\newcommand{\Deepn}{\deut($\vec{e},e^\prime \vec{n})$}
\newcommand{\Heen}{\phe($\vec{e},e^\prime n$)}
\newcommand{\Hee}{\phe($\vec{e},e^\prime$)}
\newcommand{\Ee}{\ensuremath{E_e}} 
\newcommand{\thetae}{\ensuremath{\theta_e}} 
\newcommand{\GD}{\ensuremath{G_D}}
\newcommand{\GE}{\ensuremath{G_E}}
\newcommand{\GEp}{\ensuremath{G_E^{p}}}
\newcommand{\GEpGMp}{\ensuremath{G_E^p/G_M^p}}
\newcommand{\GEn}{\ensuremath{G_E^{n}}}
\newcommand{\GM}{\ensuremath{G_M}}
\newcommand{\GMp}{\ensuremath{G_M^{p}}}
\newcommand{\GMn}{\ensuremath{G_M^{n}{}}}
\newcommand{\Q}{\ensuremath{Q^{2}{}}}
\newcommand{\updeg}{$^{o}$}
\newcommand{\etal}{et~al.}

\section{Nucleon Form Factors}

\subsection{Theory of Electron Scattering and Form Factor Measurements}

\setlength{\parindent}{0.25in}
The nucleon EMFFs are of fundamental 
importance for the understanding of the nucleon's internal structure. 
Under Lorentz invariance, spatial symmetries, and charge conservation, the most general form of the electromagnetic current inside a nucleon can be written as:

\begin{equation} 
J_{EM}^\mu = F_1(Q^2) \gamma ^\mu + \frac{\kappa}{2 M_N} F_2(Q^2) i \sigma ^{\mu \nu} q_\nu,
\end{equation}

\begin{figure}[h!]
\epsfxsize =4 cm		
\centerline{\epsfbox{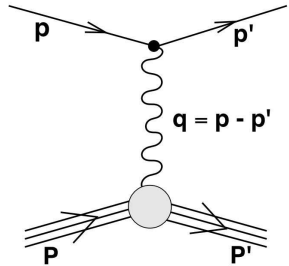}}
\caption{The Feynman diagram for the scattering of an electron with four-momentum 
$p= (E_e, \vec{p})$ through an angle $\theta_e$ off a nucleon with mass $M_N$ and four-momentum $P$. In this diagram a single virtual photon with four-momentum 
$q= p - p' = (\omega, \vec{q})$ is exchanged. The four-momenta of the scattered electron
and nucleon are $p' = (E'_e, \vec{p'})$ and $P'$, respectively.}
\label{Feyn}
\end{figure}

\setlength{\parindent}{0em}
where $F_1$ denotes the helicity non-flip Dirac form factor, $F_2$ the helicity flip Pauli form factor, $Q^2 = -q^2$, and  $\kappa$ the nucleon anomalous magnetic moment. The remaining variables are defined in Figure \ref{Feyn}. The second term, usually referred to as the Foldy contribution, 
is of relativistic origin. It is useful to introduce the isospin form-factor components, corresponding to the isoscalar ($s$) and isovector ($v$) response of the nucleon,

\begin{equation} 
F_i^s = \frac{1}{2} (F_i^p + F_i^n);~~~~~F_i^v = \frac{1}{2} (F_i^p - F_i^n);~~~~~(i=1,2).
\end{equation}

The form factors can be continued analytically into the complex plane and can be related in
different regions through a dispersion relation of the form

\begin{equation} 
F(t) = \frac{1}{\pi} \int^\infty_{t_0} \frac{\Im F(t')}{t' - t} dt',
\end{equation}

with $t = -Q^2$, $t_0 = 9 (4) M_\pi^2$ for the isoscalar (isovector) case and $M_\pi$ the pion mass.
In the positive \Q-region, called spacelike, form factors can be measured through electron scattering,
in the negative \Q-region, called timelike, form factors can only be measured through the creation or annihilation of a $N\bar{N}$-pair.

In plane-wave Born approximation, the cross section for elastic 
electron-nucleon scattering can be expressed in the Rosenbluth\cite{rose} formula as:

\begin{equation} 
\frac{d\sigma}{d\Omega} = \sigma_M[(F_1^2 + \kappa ^2 \tau 
F_2^2) + 2 \tau (F_1 + \kappa F_2)^2\tan^2(\frac{\thetae}{2})],
\end{equation}

\setlength{\parindent}{0em}
where $\tau = Q^2 /(4 M_N^2)$ and $\sigma_{M}=(\frac{\alpha_{QED} \cos \thetae /2}{2\Ee \sin ^2 \thetae/2})^2 \frac{\Ee'}{\Ee}$ is the Mott cross section for scattering off a point-like particle, with
$\alpha_{QED}$ denoting the fine-structure constant. 
The remaining variables are defined in Figure \ref{Feyn}. Hofstadter\cite{hofs} determined the values of $F_1$ and $F_2$ from measurements at different scattering angles but at the same values of $Q^2$ by drawing intersecting ellipses.   Hand, Miller and Wilson\cite{hand} pointed out that a simple algebraic separation is possible if one expresses the Rosenbuth formula in an alternate form:

\begin{equation} 
\frac{d\sigma}{d\Omega} = \sigma_M[\frac{(G^p_E)^2 + \tau (G^p_M)^2}{1 + \tau} + 2 \tau (G^p_M)^2 \tan^2(\frac{\thetae}{2})] =\frac{ \sigma_M}{\epsilon}[\tau (G^p_M)^2 + \epsilon (G^p_E)^2 ] (\frac{1}{1 + \tau}),
\label{rosen}
\end{equation}

\setlength\parindent{0em}
with $\epsilon = 1/[1 + 2(1 + \tau) \tan^2(\frac{\thetae}{2})]$ the linear polarization of the virtual photon. They further noted that $G_E$ and $G_M$ were identical to the electric and magnetic form factors, discussed earlier by Ernst, Sachs and Wali\cite{ernst}:

\begin{eqnarray} 
 G_E(Q^2) = F_1(Q^2) - \tau \kappa F_2(Q^2); & G_E^p(0) = 1; & G_E^n(0) = 0;\nonumber \\
 G_M(Q^2) = F_1(Q^2) + \kappa F_2(Q^2) ; & G_M^{p,n}(0) = \mu _{p,n},
\end{eqnarray}

\setlength\parindent{0em}
with $\mu_{p,n}$ denoting the magnetic moment of the proton and neutron, respectively. Equation \ref{rosen} illustrates that \GEp\ and \GMp\ can be determined separately by 
performing cross-section measurements at fixed \Q~as a function of $\epsilon$, over a range of 
(\thetae,\Ee) combinations (Rosenbluth separation).
Hand, Miller and Wilson further noted that in the Breit frame, which for elastic scattering is equivalent to the centre-of-mass frame, the electromagnetic current of the proton simplifies into the expression:

\begin{equation} 
J_{EM}^\mu = G_E(Q^2) \gamma ^\mu + G_M(Q^2) i \sigma ^{\mu \nu} q_\nu .
\label{current}
\end{equation}

In this reference frame the Sachs form factors can be identified with the Fourier transform 
of the nucleon charge and magnetization density distributions. 

\setlength\parindent{0.25in}
Through the mid-1990s practically all available 
proton EMFF data had been collected using the Rosenbluth separation 
technique. This experimental procedure requires an accurate 
knowledge of the electron energy and the total luminosity. In 
addition, because the \GMp\ contribution to the elastic cross section
 is weighted with \Q, data on \GEp\ suffer 
from increasing systematic uncertainties with increasing \Q-values. The 
then available world data set\cite{bost} was compared to the 
so-called dipole parametrization \GD, which corresponds to two poles with 
opposite sign close to each other in the time-like region. In coordinate space \GD\
corresponds to exponentially 
decreasing radial charge and magnetization  densities, albeit with a non-physical
discontinuity at the origin:

\begin{equation} 
 G_D = \left( \frac{\Lambda ^2} {\Lambda ^2+Q^2} \right)^2
 \mbox{~~~~~with $\Lambda$ = 0.84~GeV and $Q$ in GeV}.
\end{equation}
 
\begin{figure}[h!]
\epsfxsize =12 cm		
\centerline{\epsfbox{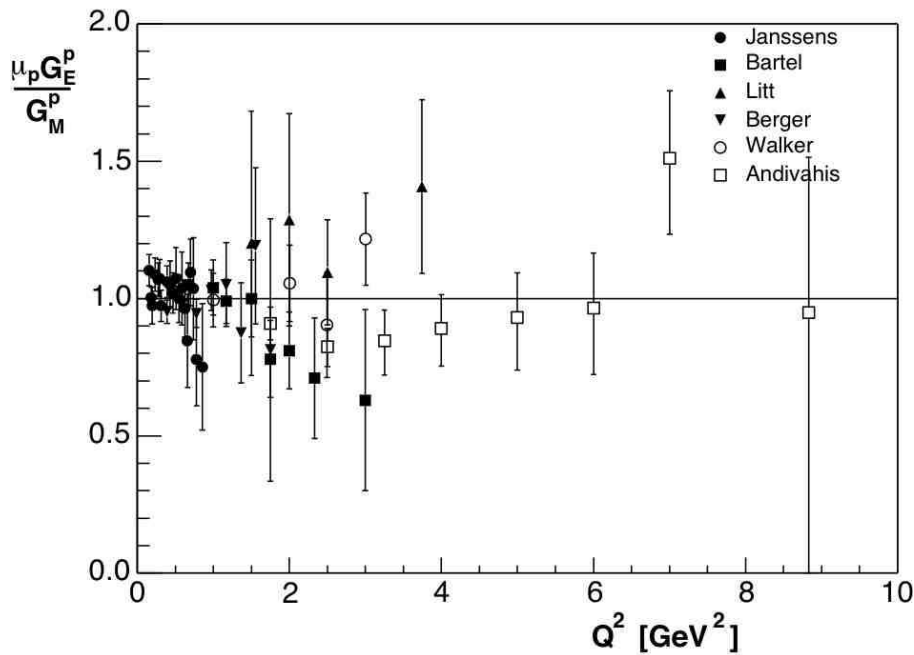}}
\caption{The ratio $\mu_p \GEpGMp$ from Rosenbluth separation.
Data are from References \cite{jans,bart,litt,berg,walk,andi}. The errors shown are the
quadratic sum of the statistical and systematic contributions.}
\label{Rosen}
\end{figure}

For \GEp, $\GMp/\mu_p$ and $\GMn/\mu_n$ the available data agreed to within 20\%
with the dipole parametrization. Both the \GEp\ and 
the $\GMp/\mu_p$ data could be fitted adequately with an 
identical parametrization. However, the limitation of the Rosenbluth 
separation was 
evident from the fact that different data sets for $\mu_p G_E^p/ G_M^p$ scattered
by up to 50\% at \Q-values larger than 1 \GeV\ (Figure \ref{Rosen}).
Although no fundamental reason has been found for the success of the dipole parametrization, it is still used as a base line for comparison of data because it takes out the largest variation with $Q^2$ and enables smaller differences to be seen.

\subsection{Instrumentation for Form Factor Measurements}

More than 40 years ago Akhiezer et al.\cite{akhi} (followed 20 years later by Arnold 
\etal\cite{arnold}) showed that the accuracy of nucleon charge  form-factor measurements 
could be increased significantly by scattering 
polarized electrons off a polarized target (or equivalently by
measuring the polarization of the recoiling proton). 
However, it took several decades before technology had sufficiently advanced
to make the first of such measurements feasible, and only in the past few years
has a large number of new data with a significantly improved accuracy become available.
The next few sections introduce the various techniques.
For \GEp\ measurements, the highest figure of merit 
at \Q-values larger than a few \GeV\ is obtained with a focal plane polarimeter.
Here, the Jacobian focusing of the recoiling proton kinematics allows one to couple
a standard magnetic spectrometer for the proton detection to a large-acceptance
non-magnetic detector for the detection of the scattered electron. For studies of
\GEn\, one needs to use a magnetic spectrometer to detect
the scattered electron in order to cleanly identify the reaction channel. As a consequence,
the figure of merit of a polarized \phe\ target is comparable to that of a neutron polarimeter. 

\subsubsection{Polarized Beam}

Various techniques are available to produce polarized electron beams, but photo-emission 
from GaAs has until now proven to be optimal\cite{aule}. A thin layer of GaAs is illuminated
by a circularly polarized laser beam of high intensity, which preferentially excites electrons 
of one helicity state to the conductance band through optical pumping. The helicity sign of the laser beam
can be flipped at a rate of tens of Hertz by changing the high voltage on a Pockels cell. The polarized 
electrons that diffuse to the photocathode surface  are then extracted by a 50-100 kV potential.
An ultra-high vacuum environment is required to minimize surface degradation of the
GaAs crystal by backstreaming ions. Initially, the use of bulk GaAs limited the maximum 
polarization to 50\% because of the degeneracy of the $P_{3/2}$ sublevels. This degeneracy 
is removed by introducing a strain in a thin layer of GaAs deposited onto a thicker layer with a 
slightly different lattice spacing. Although such strained GaAs cathodes have a significantly
lower quantum efficiency than bulk GaAs cathodes, this has been compensated by the 
development of high-intensity diode or Ti-sapphire lasers.
Polarized electron beams are now reliably available 
with a polarization close to 80\% at currents of $\ge 100~\mu$A.

The polarized electrons extracted from the GaAs surface are first pre-accelerated and longitudinally 
bunched and then injected into an accelerator. Typically, the polarization vector of the electrons is 
manipulated in a Wien filter so that the electrons are fully longitudinally polarized at the target. 
If the beam is injected into a storage ring for use with an internal target, a Siberian snake\cite{derb} is needed to compensate for the precession of the polarization.

Three processes are used to measure the beam polarization: Mott\cite{stei} scattering, M{\o}ller\cite{haug} 
scattering or Compton\cite{baylac} scattering. Any of these results in a
polarimeter with an accuracy approaching 1\%. In a Mott polarimeter the beam helicity asymmetry is measured in scattering polarized electrons off atomic nuclei. This technique is limited to electron energies below $\sim 20$ MeV and multiple scattering effects have to be estimated by taking measurements at different target foil thicknesses. In a M{\o}ller polarimeter  polarized electrons are scattered off polarized atomic electrons in a magnetized iron foil. In this technique the major uncertainties are in the corrections for atomic screening and in the foil magnetization, unless the polarizing field is strong enough to
saturate the magnetization. A potentially superior alternative\cite{chudakov} has been proposed in which the electrons are scattered off a sample of atomic hydrogen, polarized to a very high degree in an atomic beam, and trapped in a superconducting solenoid. Finally, in a Compton polarimeter the beam helicity asymmetry is measured in scattering polarized electrons off an intense beam of circularly polarized light,  produced by trapping a laser beam in a high-finesse Fabry-Perot cavity. The electron beam in a storage ring is sufficiently intense that a laser beam can be directly scattered off the electron beam without the use of an amplifying cavity. Only the last two methods, the atomic hydrogen M{\o}ller and the Compton polarimeter, have no effect on the quality of the electron beam and thus can be used continuously during an experiment.

\subsubsection{Polarized Targets}
In polarized targets for protons two different techniques are used, depending on the intensity of the
electron beam. In storage rings where the circulating beam can have an intensity of 100 mA
or more, but the material interfering with the beam has to be minimized, gaseous targets are
used, whereas in external targets solid targets can be used. Because free neutrons are not available in 
sufficient quantity, effective targets, such as deuterium or $^3$He, are necessary, and the
techniques used to polarize the deuteron are similar to those used for the proton. For $^3$He gaseous targets are used both in internal and 
external targets.

Solid polarized targets that can withstand electron beams with an intensity of up to 100 nA all use
the dynamic nuclear polarization technique\cite{crab}. A hydrogenous compound, such as 
NH$_3$ or LiD, is doped, e.g. by radiation damage, with a small concentration of free radicals. Because the occupation of the magnetic substates in the radicals follows the Boltzmann distribution, the free electrons are polarized to more than 99\% in a $\sim 5$ T magnetic field and at a $\sim 1$ K temperature. A radiofrequency (RF) field is then applied to induce transitions to states with a preferred orientation of the nuclear spin. Because the relaxation time of the electrons is much shorter than that of the nuclei, polarized nuclei are accumulated. This technique has been successful in numerous deep-inelastic lepton scattering and nucleon form-factor experiments; it has provided polarized hydrogen or deuterium targets with an average polarization of $\sim 80$\% or $\sim 30$\%, respectively. 

Internal hydrogen/deuterium targets\cite{steffens} are polarized by the atomic beam source (ABS) technique, which relies on Stern-Gerlach separation and RF transitions. First, a beam of atoms is produced in an RF dissociator through a nozzle cooled with liquid nitrogen. Then, atoms with different electron spin direction are separated through a series of permanent (or superconducting) sextupole magnets and transitions between different hyperfine states are induced by a variety of RF units. The result is a highly polarized beam  with a flux
up to $10^{17}$ atoms/s. This beam is then fed into an open-ended storage cell, which is cooled and coated to minimize recombination of the atoms bouncing off the cell walls. The circulating electron beam, passing through the
long axis of the storage cell, encounters only the flowing atoms. The polarization vector is oriented with a
set of coils, producing a field of $\sim 0.3$ T  in order to minimize depolarization by
the RF structure of the circulating electron beam. The diameter of the storage cell is determined by the
halo of the electron beam. A target thickness of $2 \times 10^{14}$ nuclei/cm$^2$ has been obtained at a vector 
polarization of more than 80\%. 

Polarized hydrogen or deuterium atoms can also be produced by spin-exchange collisions between such atoms and a small admixture of alkali atoms that have been polarized by optical pumping. The nucleus is then polarized in spin-temperature equilibrium. Although the nuclear polarization obtained in such a laser driven source (LDS) is smaller than through the ABS technique, the flux can be more than $10^{18}$ atoms/s. Moreover, an LDS offers a more compact design than an ABS. A figure of merit comparable to that of the ABS at the HERMES experiment has recently been achieved by the MIT group\cite{clasie}.

Polarized $^3$He is attractive as an effective polarized neutron target because its ground state is 
dominated by a spatially symmetric $s$-state in which the proton spins cancel, so that the spin of the $^3$He
nucleus is mainly determined by that of the neutron. Corrections for the (small) $d$-state component
and for charge-exchange contributions from the protons can be calculated accurately at \Q-values smaller than 
0.5 \GeV \cite{gola} and larger than $\sim 2$ \GeV \cite{sarg}. Direct optical pumping of $^3$He atoms is not possible because of the energy difference between the ground state and the first excited state. Instead $^3$He is polarized, either by first exciting the atoms to a metastable $2^3S_1$ state and optically pumping that state, which then transfers its polarization to the ground state by metastability-exchange collisions, or by optically pumping a small admixture of rubidium atoms, which then transfer their polarization to the $^3$He atoms through spin-exchange collisions. In internal targets only the metastability-exchange technique has been used because of the possible detrimental effects of the rubidium admixture on the storage ring environment. With beam on target, polarization values of up to 46\% at target thicknesses of $1 \times 10^{15}$ nuclei/cm$^2$ have been obtained. For external targets the spin-exchange technique\cite{alcorn} has been used to optically pump a glass target cell filled with 10 atm of $^3$He with a 
0.1\% rubidium admixture. After the spin-exchange collisions the polarized $^3$He diffuses into a 25 cm long cell which the electron beam traverses. Polarizations in excess of 40\% have been reached with beam on target. A pair of 5 mT Helmholz coils is used to orient the polarization vector, and care must be taken to minimize depolarizing magnetic field gradients. Alternatively, the metastability technique\cite{surk} has been used to polarize $^3$He under atmospheric pressure which is then compressed to a density of more than 6 atm.

\subsubsection{Recoil Polarimeters}
Focal-plane polarimeters have long been used at proton scattering facilities to measure the polarization of the
scattered proton. In such an instrument\cite{alcorn} the azimuthal angular distribution is measured of protons scattered in the focal plane of a magnetic spectrometer by an analyzer, which often consists of carbon. From this angular distribution the two polarization components transverse to the proton momentum can be derived. The analyzer is preceded by two detectors, most often wire or straw chambers, to measure the track of the incident proton; it is followed by two more detectors to track the scattered particle. The thickness of the analyzer is adjusted to the proton momentum, limiting multiple scattering while optimizing the figure of merit. In order to determine the two polarization components in the scattering plane at the target, care must be taken to accurately calculate on an event-by-event basis the precession of the proton spin in the magnetic field of the spectrometer.

Neutron polarimeters follow the same basic principle. Here, plastic scintillator material is used as an active analyzer, preceded by a veto counter to discard charged particles. This eliminates the need for the front detectors. Sets of scintillator detectors are used to measure an up-down asymmetry in the scattered neutrons, which is sensitive to a polarization component in the scattering plane, perpendicular to the neutron momentum. In modern neutron polarimeters\cite{ostr} the analyzer is preceded by a dipole magnet, with which the neutron spin can be precessed.

\subsection{Experimental Results}

\subsubsection{Proton Electric Form Factor}

In elastic electron-proton scattering a longitudinally polarized electron will transfer its 
polarization to the recoil proton. In the one-photon exchange approximation the proton 
can attain only polarization components in the scattering plane, parallel ($P_l$) and 
transverse ($P_t$) to its momentum. This can be immediately seen from the expression for the proton current in the Breit frame, which separates into components proportional to $G_E$ and $G_M$ to which reference was made in eq. \ref{current}. The ratio of the charge and magnetic form factors is
directly proportional to the ratio of these polarization components\cite{perdri}:

\begin{equation} 
\frac{\GEp}{\GMp} = - \frac{P_t}{P_l} \frac{\Ee+\Ee'}{2M} \tan(\frac{\thetae}{2}).
\end{equation}
 
The polarization-transfer technique was used for the first time by Milbrath \etal \cite{milb}
at the MIT-Bates facility. The proton form factor ratio was measured at  \Q-values of 0.38 
and 0.50 \GeV\ by scattering a 580 MeV electron beam polarized to $\sim 30$\%.
A follow-up measurement was performed at the MAMI facility\cite{posp} at a \Q-value of 0.4 \GeV.

\begin{figure}[h!]
\epsfxsize =12 cm		
\centerline{\epsfbox{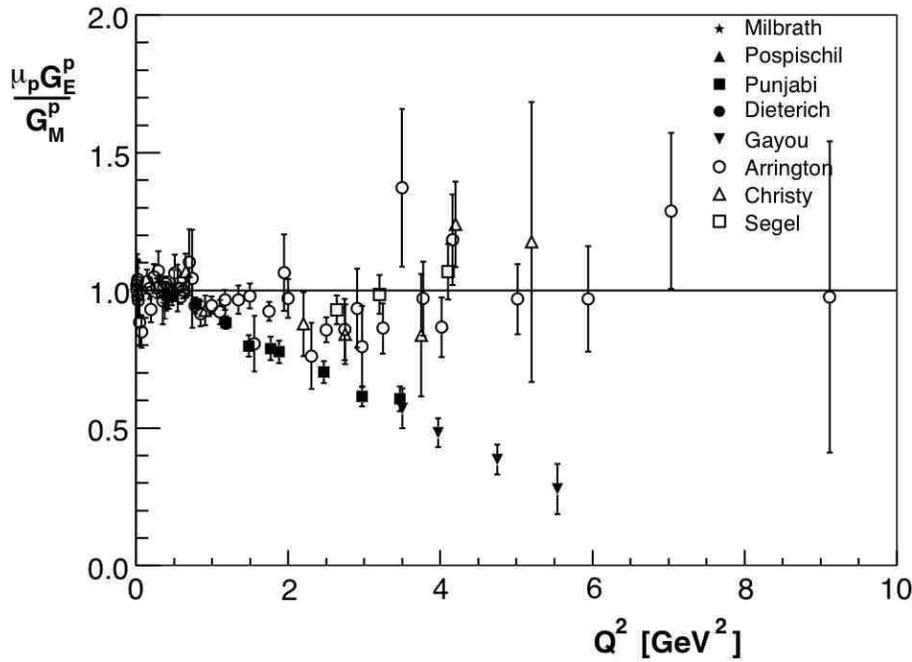}}
\caption{The ratio $\mu_p \GEpGMp$ from polarization transfer\cite{milb,posp,punj,diet,gayou},
compared to recent Rosenbluth data\cite{segel,christy} and the reanalysis by Arrington\cite{arrington1} of older SLAC data. }
\label{GEp}
\end{figure}

The greatest impact of the polarization-transfer technique was made by the two recent
experiments\cite{punj,gayou} in Hall A at Jefferson Lab, which measured the ratio \GEpGMp\ in 
a \Q-range from 0.5 to 5.6 \GeV. 
 Elastic $ep$ events were selected by detecting electrons 
 and protons in coincidence in the two identical high-resolution spectrometers.
 At the four highest \Q-values a lead-glass calorimeter was used to 
 detect the scattered electrons in order to match the proton angular acceptance.
 The polarization of the recoiling proton was determined with a 
 focal-plane polarimeter in the hadron spectrometer, consisting of 
 two pairs of straw chambers with a carbon or polyethylene analyzer 
 in between. 
The data were analyzed in bins of each of the target coordinates. 
No dependence on any of these variables was observed\cite{punj}.
Figure \ref{GEp} shows the results for the ratio $\mu_p G_E^p/ G_M^p$ . The 
most striking feature of the data is the sharp, practically linear decline 
as \Q\ increases:

\begin{equation} 
\mu_p  \frac{\GEp(\Q)}{\GMp(\Q)} = 1 - 0.13(\Q - 0.29 \rm{GeV^2}).
\end{equation}
 
Since it is known that \GMp\ closely follows the dipole 
parametrization, it follows that \GEp\ falls more rapidly with \Q\ 
than $G_{D}$. This significant fall-off of the form-factor ratio is in clear 
disagreement with the results from the Rosenbluth extraction.
Arrington\cite{arrington1} has performed a careful reanalysis of earlier Rosenbluth data.
He selected only experiments in which an adequate $\epsilon$-range
was covered with the same detector. The results (Figure \ref{GEp}) do not show the large
scatter seen in Figure \ref{Rosen}.
Recently, Christy \etal\cite{christy}~analyzed an extensive data set on elastic electron-proton 
scattering collected in Hall C at Jefferson Lab as part of experiment E99-119. The results
are evidently in good agreement with Arrington's reanalysis. 
Qattan \etal\ \cite{segel} performed a high-precision Rosenbluth extraction 
in Hall A at Jefferson Lab, designed specifically
to significantly reduce the systematic errors compared to earlier Rosenbluth
measurements. The main improvement came from detecting the recoiling
protons instead of the scattered electrons, so that the proton momentum 
and the cross section remain practically constant when one varies $\epsilon$
at a constant \Q-value. In addition, possible dependences on the beam 
current are minimized. Special care was taken in surveying the angular setting
of the identical spectrometer pair. One of the spectrometers was used as a 
luminosity monitor during an $\epsilon$ scan. The results\cite{segel} of this 
experiment, covering \Q-values from 2.6 to 4.1 \GeV, are in excellent agreement
with previous Rosenbluth results. This basically rules out the possibility that
the disagreement between Rosenbluth and polarization-transfer measurements
of the ratio \GEpGMp\ is due to an underestimate of $\epsilon$-dependent
uncertainties in the Rosenbluth measurements.

\subsubsection{Two-Photon Exchange}

\begin{figure}[h!]
\epsfxsize =8 cm	
\centerline{\epsfbox{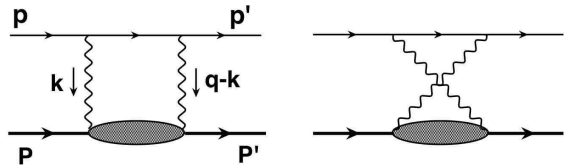}}
\caption{The Feynman diagrams depicting two-photon exchanges.}
\label{2photon}
\end{figure}

In order to resolve the discrepancy between the results for \GEpGMp\ from the 
two experimental techniques, an $\epsilon$-dependent modification of the
cross section is necessary. In two-(or more-)photon exchanges
(TPE) the nucleon undergoes a first virtual photon exchange which can lead to
an intermediate excited state and then a second one or more, finally ending back in its 
ground state (Figure \ref{2photon}). The TPE contributions to elastic electron scattering have been investigated both experimentally and theoretically for the past fifty
years. In the early days such contributions were called dispersive effects\cite{offer}. Lately, they have 
been relocated to radiative corrections in the so-called box diagram. Almost all analyses
with the Rosenbluth technique have applied radiative corrections using the formulae derived by Mo \& 
Tsai\cite{tsai} that only include the infrared divergent parts of the box diagram
(in which one of the two exchanged photons is soft). Thus, terms in which both photons are hard
(and which depend on the hadronic structure) have been ignored.

The most stringent tests of TPE on the nucleon have been carried out by measuring the
ratio of electron and positron elastic scattering off a proton. Corrections due to TPE
will have a different sign in these two reactions. Unfortunately, this (e$^+$e$^-$) data set is 
quite limited\cite{arrington2}, only extending (with poor statistics) up to a \Q-value of $\sim 5$ \GeV, whereas at \Q-values larger than $\sim 2$ \GeV\ basically all data have been measured at 
$\epsilon$-values larger than $\sim 0.85$. Other tests, also inconclusive, searched for
non-linearities in the $\epsilon$-dependence or measured the transverse (out-of-plane) 
polarization component of the recoiling proton, of which a non-zero value would be a direct 
measure of the imaginary part of the TPE amplitude.

Several studies have provided estimates of the size
of the $\epsilon$-dependent corrections necessary to resolve the discrepancy. 
Because the fall-off
of the form-factor ratio is linear with \Q, and the Rosenbluth formula also shows
a linear dependence of the form-factor ratio (squared) with \Q\ through the $\tau$-term,
a \Q-independent correction linear in $\epsilon$ would cancel the 
disagreement. An additional constraint that any $\epsilon$-dependent modification
must satisfy, is the (e$^+$e$^-$) data set. 
Guichon \& Vanderhaeghen\cite{guichon} introduced a general form of a TPE 
contribution from the so-called box diagram in radiative corrections into the
amplitude for elastic electron-proton scattering.
%
%
This resulted in the following modification of the Rosenbluth
expression:

\begin{equation} 
d \sigma \propto {\tau + \epsilon \frac{\tilde{G_E}^2}{\tilde{G_M}^2} + 2 \epsilon 
(\tau + \frac {\tilde{G_E}}{\tilde{G_M}})Y_{2 \gamma} },
\end{equation}
 
\setlength{\parindent}{0em}
where $Y_{2 \gamma} = \Re{\frac{\nu \tilde{F_3}}{M^2 \tilde{G_M}}}$ and $\tilde{G_M}$,  $\tilde{F_2}$ and $\tilde{F_3}$ are equal to $G_M$,  $F_2$ and 0, respectively, in the Born approximation. $Y_{2 \gamma}$ and 
the "two-photon" form factors $\tilde{G_E}$ and $\tilde{G_M}$ were fitted\cite{guichon} to the Rosenbluth 
and polarization transfer data sets. This resulted in a value of $\sim$ 0.03
for $Y_{2 \gamma}$ with very little $\epsilon$- or \Q-dependence. 

Arrington\cite{arrington3} performed a fit to the complete data set, investigating two different
modifications to the cross section with a \Q-independent linear $\epsilon$-dependence of 6 \%
over the full $\epsilon$-range.
Both modifications have the same $\epsilon$-dependence, but one does not 
modify the cross section at small values of $\epsilon$, whereas the other leaves the 
cross section unchanged at large values of  $\epsilon$. He found that the second
gave a much better description of the complete data set. Moreover, it
was in good agreement with the data set for the ratio of electron-proton and 
positron-proton elastic scattering.

Blunden \etal\cite{blunden} carried out the first calculation of the elastic contribution from 
TPE effects, albeit with a simple monopole \Q-dependence of the hadronic  form
factors: $G(Q^2)= \Lambda ^2 / (Q^2 + \Lambda ^2)$. They obtained a practically 
\Q-independent correction factor with a linear $\epsilon$-dependence that vanishes
at forward angles ($\epsilon = 1$). However, the size of the correction only resolves about
half of the discrepancy. A later calculation(W. Melnitchouk, private communication) which used a more realistic form factor behavior, resolved up to 80\% of the discrepancy.

A different approach was used by Chen et al.\cite{afanasev}, who related the elastic
electron-nucleon scattering to the scattering off a parton in a nucleon through generalized
parton distributions. TPE effects in the lepton-quark scattering process
are calculated in the hard-scattering amplitudes. The handbag formalism of the generalized
parton distributions is
extended in an unfactorized framework in which the $x$-dependence is retained in the 
scattering amplitude. Finally, a valence model is used for the generalized
parton distributions. The results for the
TPE contribution fully reconcile the Rosenbluth and the polarization-transfer
data and retain agreement with positron-scattering data.

Hence, it is becoming more and more likely that TPE processes have to 
be taken into account in the analysis of Rosenbluth data and that they will affect polarization-transfer
data only at the few percent level. Of course, further effort is needed to investigate the 
model-dependence of the TPE calculations. Experimental confirmation of TPE effects will be
difficult, but certainly should be continued. The most direct test would be a measurement
of the positron-proton and electron-proton scattering cross-section ratio at small
$\epsilon$-values and \Q-values above 2 \GeV. Positron beams available at
storage rings are too low in either energy or intensity, but a measurement in the CLAS detector at Jefferson Lab, a more promising venue, has been proposed\cite{broo2}.
A measurement of the beam or target single-spin asymmetry normal to the scattering plane,
which directly accesses the imaginary part of the box diagrams, would provide a
sensitive test of TPE calculations.
Also, real and virtual Compton scattering data can provide additional
constraints on calculations of TPE effects in elastic scattering.
Rosenbluth analyses have so far been restricted to simple PWBA, Coulomb distortion effects
should certainly be included too.
Additional efforts should be extended to studies of TPE effects in other 
longitudinal-transverse separations, such as proton knock-out and 
deep-inelastic scattering (DIS) experiments. 

\subsubsection{Proton Magnetic Form Factor}

\begin{figure}[h!]
\epsfxsize =12 cm	
\centerline{\epsfbox{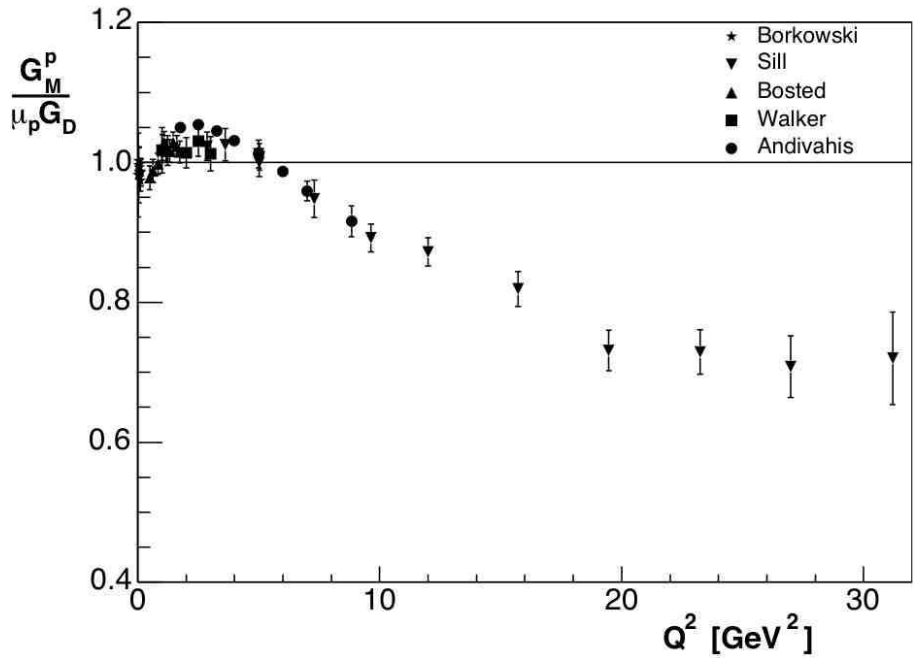}}
\caption{The proton magnetic form factor \GMp, in 
units of $\mu_p G_{D}$, as a function of \Q. Data are from References \cite{bork,bost3,sill,walk,andi}.}
\label{GMp}
\end{figure}

An extensive data set\cite{bork} with a good accuracy is available up to a \Q-value of more than 30 \GeV\ from
unpolarized cross-section measurements (Figure  \ref{GMp}). Because \GMp\ dominates in a Rosenbluth extraction at larger \Q-values, the \GMp\ data have only a minor sensitivity to the discrepancy between the
Rosenbluth extraction and the polarization-transfer technique. Brash \etal\cite{brash} have shown 
that the \GMp\ data must be renormalized upwards by $\sim$ 2\% if one assumes the
polarization-transfer data to be correct.
 
\subsubsection{Neutron Magnetic Form Factor}

Early data on \GMn\ were extracted from inclusive quasi-elastic scattering off the deuteron.
However, modeling of the deuteron wave function, required to subtract the contribution
from the proton, resulted in sizable systematic uncertainties. A significant break-through 
was made by measuring the ratio of quasi-elastic neutron and 
proton knock-out from a deuterium target. This method has little  
sensitivity to nuclear binding effects and to fluctuations in the 
luminosity and detector acceptance. The basic set-up used in all 
such measurements is very similar: the electron is detected in 
a magnetic spectrometer with coincident neutron/proton detection 
in a large scintillator array. The main technical difficulty in such 
a ratio measurement is the absolute determination of the neutron 
detection efficiency. Such measurements have been pioneered for \Q-values smaller
than 1 \GeV\ at Mainz\cite{ankl1, ankl2,kubo} and Bonn\cite{brui}. 
The Mainz  \GMn\ data are 8\%-10\% lower than those from Bonn, at variance with the 
quoted uncertainty of $\sim$2\%. This discrepancy would require a 
16\%-20\% error in the detector efficiency.

A study of \GMn\ at \Q-values up to 5 \GeV\ has recently been completed in Hall B 
by measuring the neutron/proton quasi-elastic cross-section ratio using the CLAS detector\cite{broo}.
A hydrogen target was in the beam simultaneously with the deuterium target.
This made it possible to measure the neutron detection efficiency by tagging neutrons in 
exclusive reactions on the hydrogen target. Preliminary results\cite{broo} indicate that
\GMn\ is within 10\% of \GD\ over the full \Q-range of the experiment (0.5-4.8 \GeV).

Inclusive quasi-elastic scattering of polarized electrons 
off a polarized $^3$He target offers an alternative method to determine \GMn\ through a
measurement of the beam asymmetry\cite{donn}

\begin{equation} 
A = - \frac{(\cos \theta^* v_{T'}R_{T'}+2 \sin \theta^* \cos \theta^*  v_{TL'}R_{TL'})}
{ v_L R_L + v_T R_T},
\end{equation}

\setlength{\parindent}{0em}
where $\theta^*$ and $\phi^*$ are the polar and azimuthal target spin angles with respect to $\vec{q}$, $R_i$ denote various nucleon response functions, and  $v_i$ the corresponding kinematic factors. By orienting the target polarization parallel to $\vec{q}$,
one measure $R_{T\prime}$, which in quasi-elastic kinematics is dominantly sensitive to $(\GMn)^2$.
For the extraction of \GMn\ corrections for the nuclear medium\cite{gola} are necessary to take into account effects of final-state interactions and meson-exchange currents. 
The first such measurement was carried out at Bates\cite{gao}.
Recently, this technique was used to measure \GMn\ in Hall A at 
Jefferson Lab in a \Q-range from 0.1 to 0.6 \GeV\cite{xu1}. This experiment provided 
an independent, accurate measurement of  \GMn\ at \Q-values of 0.1 and 
0.2 \GeV, in excellent agreement with the Mainz data. At the higher 
\Q-values \GMn\ could be extracted\cite{xu2} in  plane wave impulse approximation, 
since final-state interaction effects are expected to decrease with increasing \Q. 

\begin{figure}[h!]
\epsfxsize =12 cm	
\centerline{\epsfbox{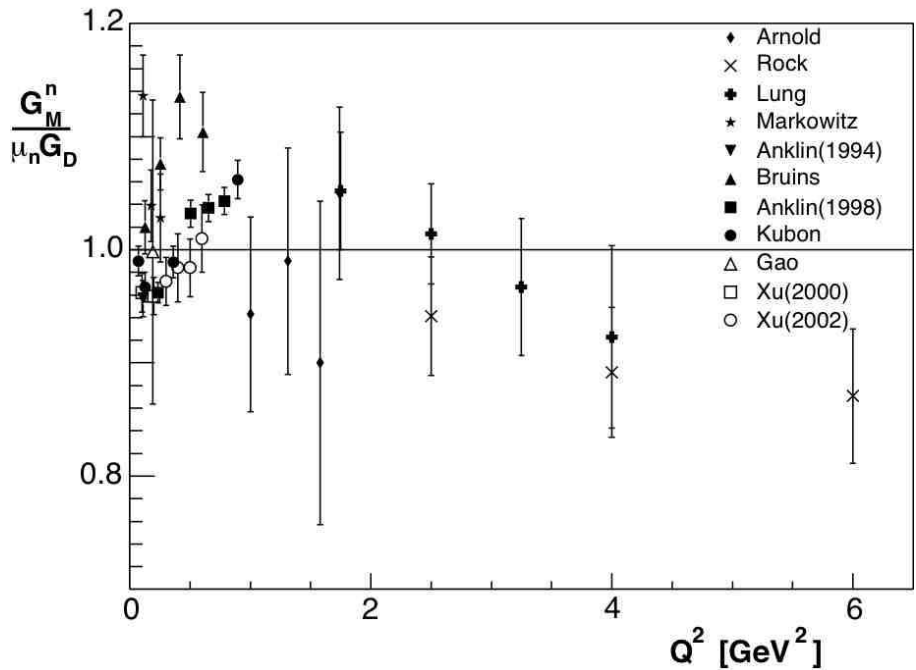}}
\caption{The neutron magnetic form factor \GMn, in 
units of $\mu_n G_{D}$, as a function of \Q. Results from \phe~are indicated by open symbols.
Data are from References \cite{arno,rock,lung,mark,ankl1,brui,ankl2,kubo,gao,xu1,xu2}.}
\label{GMn}
\end{figure}

Figure \ref{GMn} shows the results of all completed \GMn~experiments. 

\subsubsection{Neutron Electric Form Factor}

Analogously to \GMn, early \GEn-experiments used (quasi-)elastic scattering off the deuteron to extract the longitudinal deuteron response function. Due to the smallness of \GEn, the use of
different nucleon-nucleon potentials resulted in a 100\% spread in the resulting 
\GEn\ values\cite{plat}. In the past decade a series of double-polarization measurements of
neutron knock-out from a polarized $^2$H or $^3$He target have 
provided accurate data on \GEn. The ratio of the beam-target 
asymmetry with the target polarization perpendicular and 
parallel to the momentum transfer is directly proportional to 
the ratio of the electric and magnetic form factors,

\begin{equation} 
 \frac{\GEn}{\GMn} = - \frac{P_x}{P_z} \frac{\Ee + \Ee'}{2M} \tan(\frac{\thetae}{2}),
\end{equation}
 
\setlength{\parindent}{0em}
where $P_x$ and $P_z$ denote the polarization component perpendicular and parallel to $\vec{q}$.
A similar result is obtained with an unpolarized deuteron target when one 
measures the polarization of the knocked-out neutron as a function of the angle over 
which the neutron spin is precessed with a dipole magnet:

\begin{equation} 
 \frac{\GEn}{\GMn} = -  \tan(\delta) \sqrt{\frac{\tau(1 + \epsilon)}{2 \epsilon}};
\end{equation}
 
\setlength{\parindent}{0em}
here, $\delta$ denotes the precession angle where the measured asymmetry is zero.

\begin{figure}[h!]
\epsfxsize =12 cm	
\centerline{\epsfbox{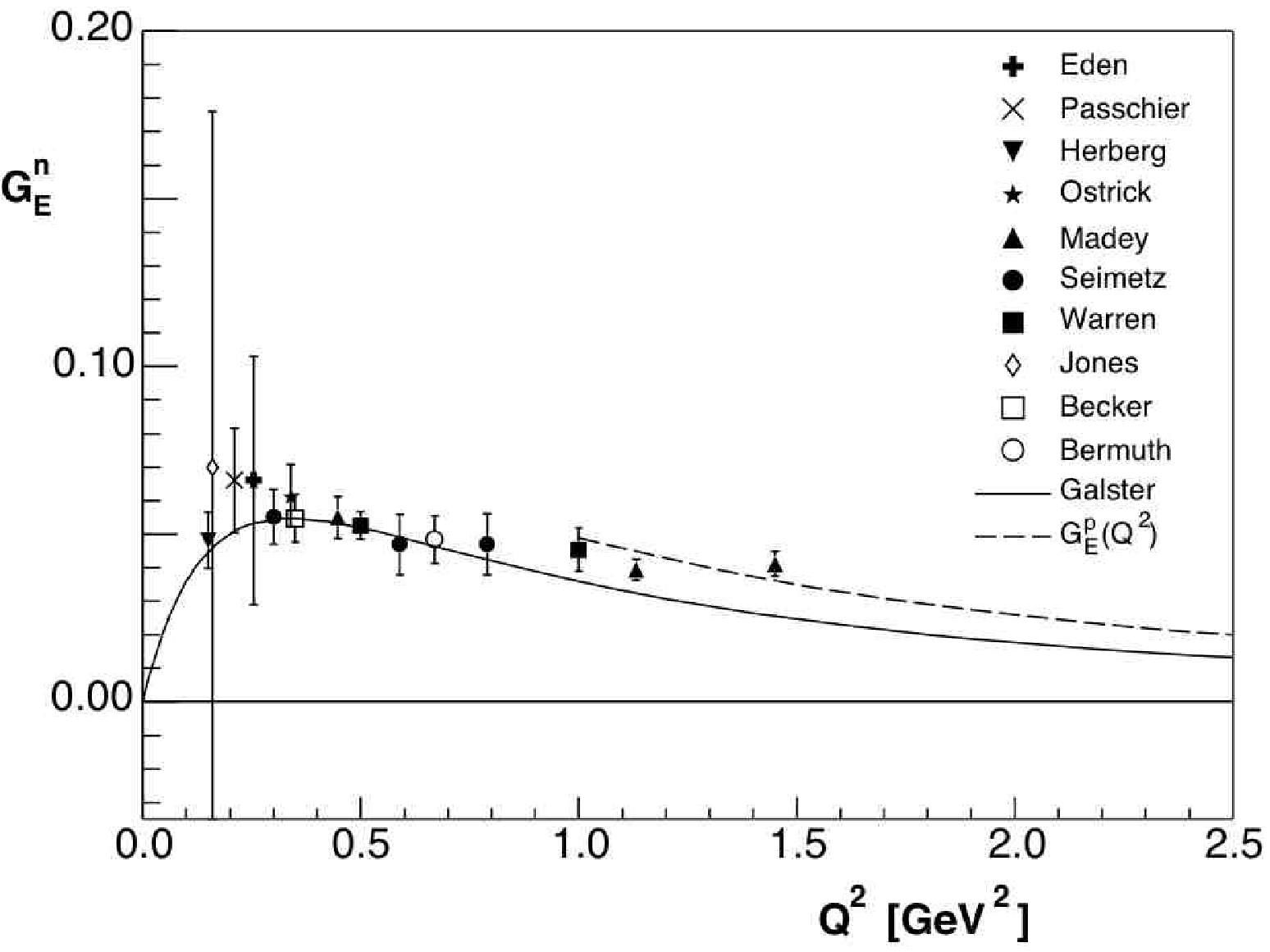}}
\caption{The neutron electric form factor \GEn~as a function of \Q.
Results from \phe~are indicated by open symbols.
Data are from References \cite{eden,pass,herb,ostr,made,seim,warr,jone,beck,gola,berm}.  The full curve shows the Galster\cite{gals}
parametrization; the dashed curve represents the \Q-behavior of \GEp.}
\label{GEn}
\end{figure}

Again, the first such measurements were carried out at Bates, both with a polarized
\phe\ target and with a neutron polarimeter. Figure \ref{GEn}  shows  results obtained through 
all three reactions  \pDeen,  \Deepn\ and  \Heen. At low \Q-values 
corrections for nuclear medium and rescattering effects can 
be sizeable: 65\% for \deut\ at 0.15 \GeV\ and 50\% for 
\he\ at 0.35 \GeV. These corrections are expected to 
decrease significantly with increasing $Q$, although no reliable 
calculations are presently available for \he\ above 0.5 \GeV. 
There is excellent agreement between the results from the different techniques. Moreover, medium effects
have clearly become negligible at $\sim$ 0.7 \GeV, even for \he. The latest data from Hall C at Jefferson Lab, using either a polarimeter or a polarized
 target \cite{made, warr},  extend up to \Q\ $\approx$ 1.5 \GeV\ with an overall 
accuracy of $\sim$10\%, in mutual agreement. From $\sim$ 1 \GeV\ onwards  \GEn\ appears
to exhibit a \Q-behavior similar to that of \GEp.
Schiavilla \& Sick\cite{schi} have extracted \GEn\ from available data on the deuteron quadrupole
form factor $F_{C2}(Q^2)$ with a much smaller sensitivity to the nucleon-nucleon potential
than from inclusive (quasi-)elastic scattering. The 30-years-old Galster parametrization\cite{gals} continues to
provide a fortuitously good description of the data.

\subsubsection{Timelike Form Factors}

\begin{figure}[h!]
\epsfxsize =10 cm	
\centerline{\epsfbox{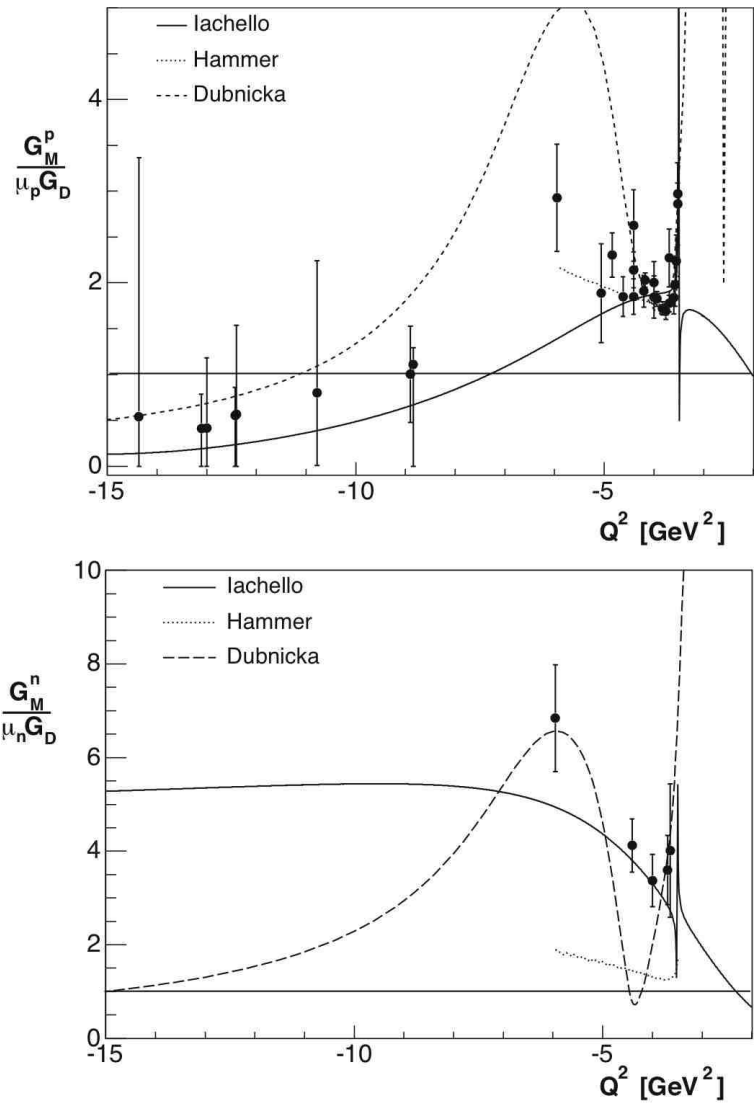}}
\caption{The magnetic form factors (divided by \GD) in the time-like region as a function of \Q, compared to the calculations by Iachello\cite{iach2}, Hammer\cite{hamm} and Dubnicka\cite{dubn}. See Reference \cite{iach2} for the references to the experimental data.}
\label{timelike}
\end{figure}

In the timelike region EMFF measurements have been made at electron-positron storage rings or by studying the inverse reaction (only for the proton form factors), antiproton annihilation on a hydrogen target.
The rather limited data set on timelike form factors is shown in Figure \ref{timelike}. The quality of the data does not allow a separation of the charge and magnetic form factors; \GM\ has been extracted from the data using the \GE-values calculated by Iachello \& Wan\cite{iach2}. Clearly \GD\, which gives a very good description of the spacelike magnetic form factors, does not describe the data in the timelike region, at least from threshold down to -6 \GeV. Iachello \& Wan\cite{iach2}, Hammer \etal\cite{hamm} and Dubnicka \etal\cite{dubn} have carried out an analytic continuation of their VMD calculations (section \ref{calc}).
Iachello's model provides a consistent description of the magnetic form factors in the timelike region.
An extension of the data set in the timelike region and of theoretical efforts to obtain a consistent description of all EMFFs in both the space- and timelike regions is highly desirable.

\subsubsection{Experimental Review and Outlook}

In recent years highly accurate data on the nucleon EMFFs have become available from various facilities
around the world, made possible by the development of high luminosity and novel polarization
techniques. These have established some general trends in the \Q-behavior of the four EMFFs. 
The two magnetic form factors \GMp\ and \GMn\ are close to identical, following \GD\ to within 10\%
at least up to 5 \GeV, with a shallow minimum at $\sim 0.25$ \GeV\ and crossing \GD\ at $\sim 0.7$ \GeV.
\GEpGMp\ drops linearly with \Q\, and \GEn\ appears to drop from $\sim 1$ \GeV\ onwards at the same rate as \GEp.
Measurements that extend to higher \Q-values and offer improved accuracy at
lower \Q-values, will become available in the near future. In Hall C at Jefferson Lab Perdrisat et al.\cite{perd} 
will extend the measurements of \GEpGMp\ to 9 \GeV\ with a new polarimeter and 
large-acceptance lead-glass calorimeter.
Wojtsekhowski et al.\cite{wojt} will 
measure \GEn\ in Hall A at \Q-values of 2.4 and 3.4 \GeV\ using the \Heen\ reaction with a 
100 msr electron  spectrometer. 
The Bates Large Acceptance Spectrometer Toroid facility (BLAST, http://www.mitbates.mit.edu) at MIT with a polarized hydrogen and deuteron target internal to a storage ring will  provide highly accurate 
data on \GEp\ and \GEn\ in a \Q-range from 0.1 to 0.8 \GeV. Gao et al.\cite{gao2} have shown that
the proton charge radius can be measured with unprecedented precision by measuring the ratio of
asymmetries in the two sectors of the BLAST detector. Thus, within a couple of years \GEn\ data with an accuracy of 10\% or better will be 
available up to a \Q-value of 3.4 \GeV. Once the upgrade to 12 GeV\cite{12gev}  has been implemented
at Jefferson Lab, it will be possible to extend the data set on \GEp\ and \GMn\ to
14 \GeV\ and on \GEn\ to 8 \GeV.

The charge and magnetization rms radii are related to the slope of the form factor at \Q = 0:

\begin{eqnarray} 
<r_E^2> = \int \rho(r) r^4 dr = \left . -6 \frac{dG(Q^2)}{dQ^2} \right |_{Q^2 = 0} \nonumber \\
<r_M^2> = \int \mu(r) r^4 dr = \left . -\frac{6}{\mu} \frac{dG(Q^2)}{dQ^2} \right |_{Q^2 = 0},
\end{eqnarray}
 
with $\rho$(r) ($\mu$(r)) denoting the radial charge (magnetization) distribution.  Table \ref{radii} lists the results. For an accurate extraction of the radius Sick\cite{sick}
 has shown that it is necessary to take into account Coulomb distortion effects and higher moments of the
radial distribution. His result for the proton charge radius is in excellent agreement with the most recent three-loop QED calculation\cite{meln} of the hydrogen Lamb shift. Within error bars the rms radii for the proton 
charge and magnetization distribution and for the neutron magnetization distribution are
equal. The value for the neutron charge radius was obtained by measuring the
transmission of low-energy neutrons through liquid $^{208}$Pb and $^{209}$Bi. The Foldy term 
$\frac{3}{2} \frac{\kappa}{M_n^2} = -0.126$ fm$^2$ is close to the value of the neutron charge radius.
Isgur\cite{isgur} showed that the Foldy term is canceled by a first-order relativistic correction, which implies that the measured value of the neutron charge radius is indeed dominated by its internal structure.

\begin{table}[h!]
\begin{center}
\caption{Values for the nucleon charge and magnetization radii}
\label{radii}
\begin{tabular}{ccc}
\hline \hline
\textbf{Observable} & \textbf{value $\pm$ error} & \textbf{Reference}  \\
\hline
$<(r_E^p)^2>^{1/2}$ & 0.895 $\pm$ 0.018 fm & \cite{sick} \\
$<(r_M^p)^2>^{1/2}$ & 0.855 $\pm$ 0.035 fm & \cite{sick} \\
$<(r_E^n)^2>$ & - 0.119 $\pm$ 0.003 fm$^2$ & \cite{kope} \\
$<(r_M^n)^2>^{1/2}$ & 0.87 $\pm$ 0.01 fm & \cite{kubo} \\
\hline
\end{tabular}
\end{center}
\end{table}

In the Breit frame the nucleon form factors can be written as Fourier transforms of their charge
and magnetization distributions. However, if the wavelength of the probe is larger than the Compton wavelength of the nucleon, i.e. if $| Q | \ge M_N$, the form factors are not solely determined by the internal structure of the nucleon. Then, they also contain dynamical effects due to relativistic boosts and consequently the physical interpretation of the form factors becomes complicated. Recently, Kelly\cite{kelly} has extracted spatial nucleon densities from the 
available form factor data. He selected a model for the Lorentz contraction of the Breit frame in which the asymptotic behavior of the form factors conformed to perturbative quantum chromo-dynamics (pQCD) scaling at large \Q-values and expanded the densities in a complete set of radial
basis functions, with constraints at large radii. The neutron and proton magnetization densities
are found to be quite similar, narrower than the proton charge density. He reports a neutron charge density with a positive core surrounded by a negative surface charge, peaking at just below 1 fm, which he attributes to a negative pion cloud. Alternatively, he extracts the radial distributions of the $u$ and $d$ quarks which both show a secondary lobe which he interprets as an indication of an orbital angular momentum (OAM) component in the quark distributions.
Friedrich \& Walcher\cite{fried} observe as a feature common to all EMFFs a bump/dip at $Q \approx$ 0.5 GeV with a width of $\sim$ 0.2 GeV. 
A fit to all four EMFFs was performed, assuming a dipole behaviour for the form factors of the constituent quarks and an $l = 1$ harmonic oscillator behaviour for that of the pion cloud. They then transformed their results to coordinate space, neglecting the Lorentz boost, where they find that the pion cloud peaks at a radius of $\sim$ 1.3 fm, slightly larger than Kelly did, close to the Compton wavelength of the pion. Hammer et al.\cite{hamm2} argue from general principles that the pion cloud should peak much more inside the nucleon, at $\sim$ 0.3 fm. However, they assign the full $N\bar{N}$2$\pi$ continuum to the pion cloud which includes different contributions than just the one-pion loop that Kelly (and Friedrich \& Walcher) assign to the pion cloud. The structure at $\sim$ 0.5 GeV, common to all EMFFs, is at such a small \Q-value that its transformation to coordinate space should be straightforward.

\subsection{Model Calculations}
\label{calc}

The recent production of very accurate EMFF data, especially the surprising \GEp\ data from polarization transfer, has prompted the theoretical community to intensify their investigation of nucleon structure. Space limitations compel us to focus on only a few highlights. The interested reader is encouraged to read the original publications; the review by Thomas \& Weise\cite{thomas} is an excellent introduction.

\begin{figure}[h!]
\epsfxsize =13 cm	
\centerline{\epsfbox{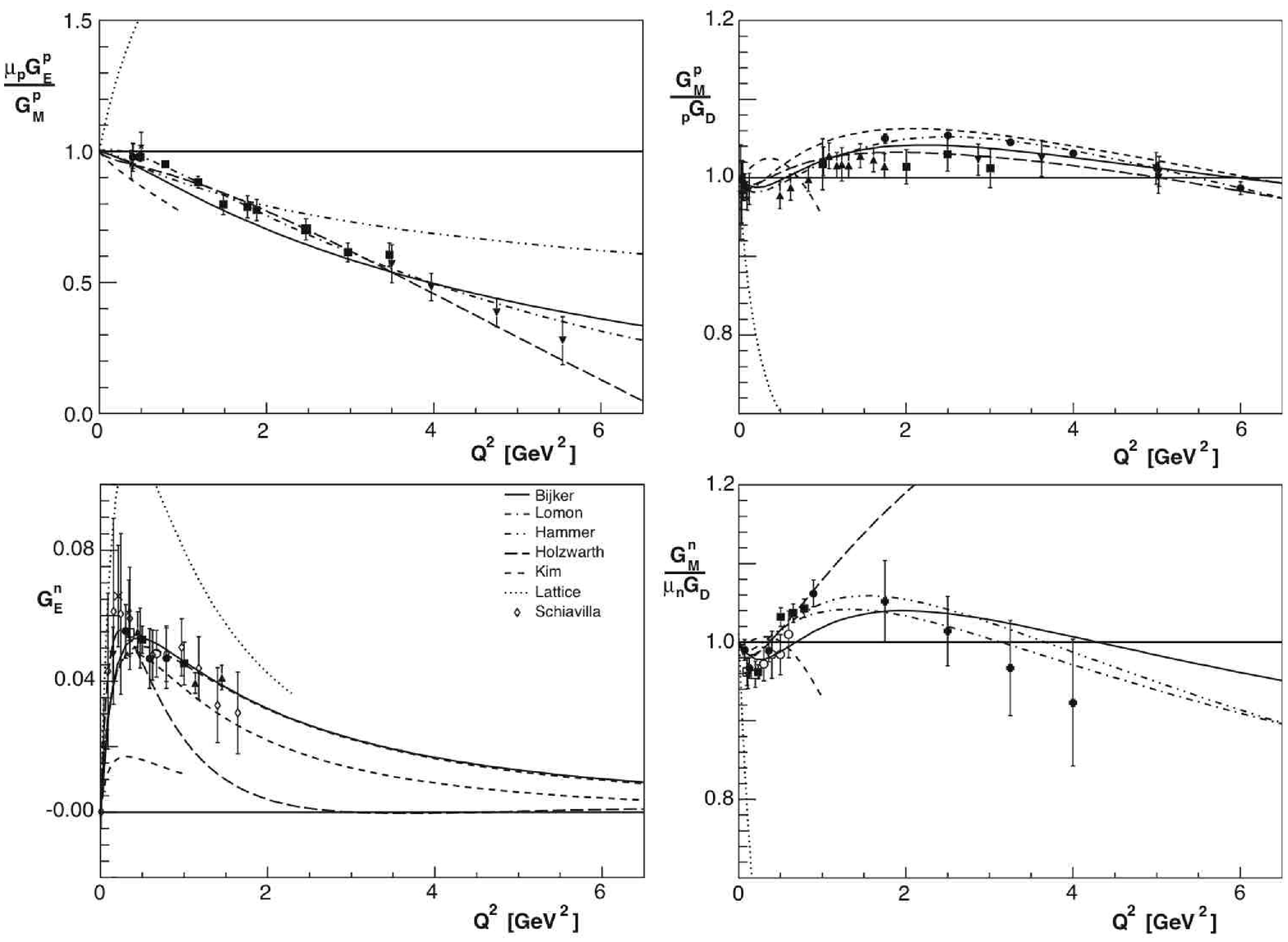}}
\caption{Comparison of various calculations with available EMFF data, indicated by the same symbols as in Figures \ref{GEp}, \ref{GMp}, \ref{GMn} and \ref{GEn}. For \GEp\ only polarization-transfer data are shown. Not shown are the data for \GEn\ of References \cite{jone,eden} and the data for \GMn\ of
References \cite{rock,arno,mark,gao,brui}. For \GEn\ the results of Schiavilla \& Sick\cite{schi} have been added.
The calculations shown are from References \cite{iach3,lomo,hamm,holz,goek,ashl}. Where applicable, the calculations have been normalized to the calculated values of $\mu_{p,n}$.}
\label{Chiral}
\end{figure}

The $u$-, $d$- and $s$-quarks are the main building blocks of the nucleon in the kinematic domain
relevant to this review. Its basic structure involves the three lightest vector mesons ($\rho$, $\omega$ and $\phi$) which have the same quantum numbers as the photon. Consequently, one should expect these vector mesons to play an important role
in the interaction of the photon with a nucleon. The first EMFF models were based on this principle, called vector meson dominance (VMD), in which one assumes that the virtual photon 
- after becoming a quark-antiquark pair - couples to the nucleon 
as a vector meson. The EMFFs can then be expressed in terms of coupling 
strengths between the virtual photon and the vector meson and between 
the vector meson and the nucleon, summing over all possible vector mesons. 
 In the scattering amplitude a bare-nucleon form factor is multiplied by the
amplitude of the photon interaction with the vector meson. With this model Iachello \etal\cite{iach} predicted a linear drop of the proton form factor ratio, similar to that measured by polarization transfer, more than 20 years before the data became available. Gari \& Kr\"{u}mpelmann\cite{gari} extended the VMD model to conform with pQCD scaling at large \Q-values. The VMD picture is not complete, as becomes obvious from the fact that the
Pauli isovector form factor $F_2^V$ is much larger than the isoscalar one $F_2^S$. An improved description requires the inclusion of the isovector $\pi \pi$ channel through dispersion relations\cite{hohl,merg}. By adding more parameters, such as the width of the $\rho$-meson and the masses of heavier vector mesons\cite{lomo}, the VMD models succeeded in describing new EMFF data as they became available, but with little predictive power. Figure \ref{Chiral} confirms that Lomon's calculations provide an excellent description of all EMFF data. Bijker \& Iachello\cite{iach3} have extended the original calculations by also including a meson-cloud contribution in $F_2$, but still taking only two isoscalar and one isovector poles into account. The intrinsic structure of the nucleon is estimated to have an rms radius of $\sim$ 0.34 fm. These new calculations are in good agreement with all EMFF data, except for $G_M^n$ at low $Q^2$-values. The most recent dispersion-theoretical analysis\cite{hamm}, using four isoscalar and three isovector mesons, results in an excellent description of \GMp\ and \GMn, but only reasonably describes \GEp\ and \GEn.
Subsequent studies\cite{fuch} have further developed this combined approach to include chiral perturbation theory. However, such models can only be used at small \Q-values, $\le 0.4$ \GeV.

Many recent theoretical studies of the EMFFs have applied various forms of a relativistic constituent quark
model (RCQM). Nucleons are assumed to be composed of three constituent quarks, which are quasi-particles where all degrees of freedom associated with the gluons and $q \bar{q}$ pairs are parametrized by an effective mass. Because the momentum transfer can be several times the nucleon mass, the constituent quarks require a relativistic quantum-mechanical treatment. Three possibilities exist for such a treatment: the instant form, where the interaction is present in the time component of the four-momentum and in the Lorentz boost;  the point form, where all components of the four-momentum operator depend on the interaction; and the light-front form, where the interaction appears in one component of the four-momentum and in the transverse rotations. In each of these forms the Poincar\'e invariance can be broken in the number of constituents (by the creation of $q\bar{q}$ pairs) or by the use of approximate current operators. Although most of these calculations correctly describe the EMFF behaviour at large \Q-values, effective degrees of freedom, such as a pion cloud and/or a finite size of the constituent quarks, are introduced to correctly describe the behaviour at lower \Q-values.

\begin{figure}[h!]
\epsfxsize =13 cm		
\centerline{\epsfbox{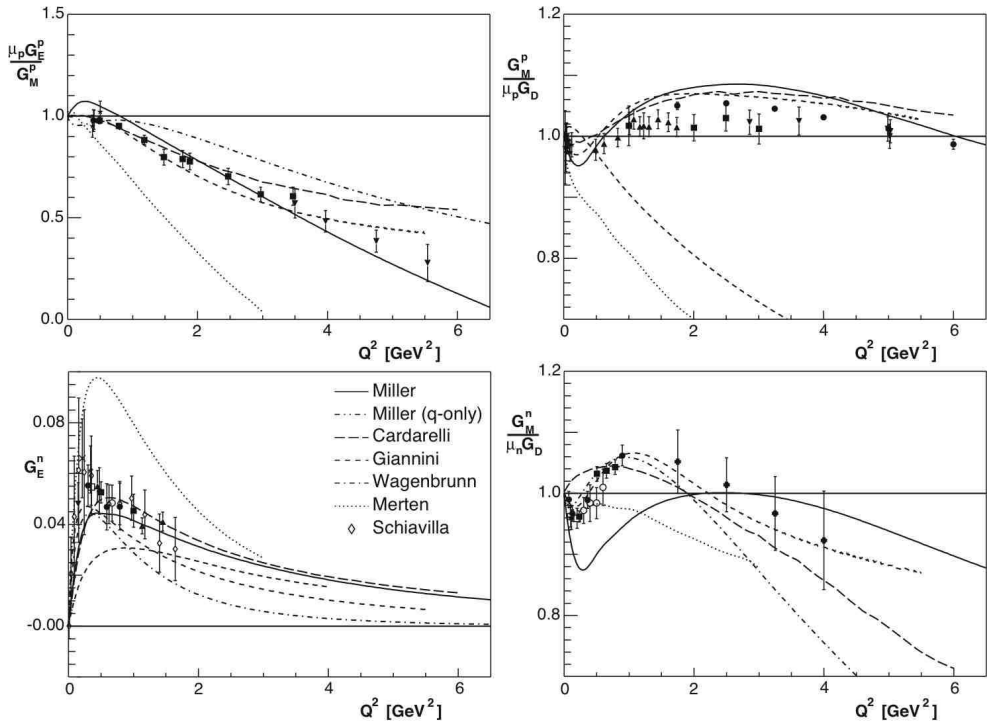}}
\caption{Comparison of various RCQM calculations with available EMFF data, similarl to the comparison in Figure \ref{Chiral}.
The calculations shown are from References \cite{mill1,simu,gian,wage,mets}. Miller (q-only) denotes a calculation by Miller\cite{mill1} in which the pion cloud has been suppressed.}
\label{RCQM}
\end{figure}

Miller\cite{mill1} uses an extension of the cloudy bag model\cite{theb}, three relativistically moving (in light-front kinematics) constituent quarks, surrounded by a pion cloud. He chose a spatial wave function, as derived by Schlumpf\cite{schl}, whose parameters (and those of the pion cloud) are chosen to describe the magnetic moments, the neutron charge radius, and the EMFF behavior at large \Q-values.  Cardarelli \& Simula\cite{simu} also use light-front kinematics, but they calculate the nucleon wave function by solving the three-quark Hamiltonian in the Isgur-Capstick one-gluon-exchange potential. In order to get good agreement with the EMFF data they
introduce a finite size of the constituent quarks in agreement\cite{petr} with recent DIS data.
The results of Wagenbrunn \etal\cite{wage} are calculated in a covariant manner
in the point-form spectator approximation (PFSA). In addition to a linear confinement, the quark-quark interaction is based on Goldstone-boson exchange dynamics. The PFSA current is effectively a three-body operator (in the case of the nucleon as a three-quark system) because of its relativistic nature. It is still incomplete but it leads to surprisingly good results for the electric radii and magnetic moments of the other light and strange baryon ground states beyond the nucleon. Although Desplanques and Theussl\cite{desp} have criticized the use of the point form in its introduction of two-body currents in the form of a neutral boson exchange, Coester and Riska\cite{coes} obtain a reasonable representation of empirical form factors in this frame. 
Giannini \etal\cite{gian} have explicitly introduced a three-quark interaction in the form of a gluon-gluon interaction in a hypercentral model, which successfully describes various static baryon properties. Relativistic effects are included by boosting the three quark states to the Breit frame and by introducing a relativistic quark current.
All previously described RCQM calculations used a non-relativistic treatment of the quark dynamics, supplemented by a relativistic calculation of the electromagnetic current matrix elements. Merten \etal\cite{mets} have solved the Bethe-Salpeter equation with instantaneous forces, inherently respecting relativistic covariance. In addition to a linear confinement potential, they used an effective flavor-dependent two-body interaction. For static properties this approach yields results\cite{caut} similar to those obtained by Wagenbrunn et al.\cite{wage}.
The results of these five calculations are compared to the EMFF data in Figure \ref{RCQM}. The calculations of Miller do well for all EMFFs, except for \GMn\ at low \Q-values. Those of Cardarelli \& Simula, Giannini \etal\ and Wagenbrunn \etal\ are in reasonable agreement with the data, except for that of Wagenbrunn \etal\ for \GMp, while the results of Merten \etal\ provide the poorest description of the data.

Before the Jefferson Lab polarization transfer data on $G_E^p/G_M^p$ became available Holzwarth\cite{holz} predicted a linear drop in a chiral soliton model. In such a model the quarks are bound in a nucleon by their interaction with chiral fields. In the bare version quarks are eliminated and the nucleon becomes a skyrmion with a spatial extension, but the Skyrme model provided an inadequate description of the EMFF data. Holzwarth's extension introduced one vector-meson propagator for both isospin channnels in the Lagrangian and a relativistic boost to the Breit frame. His later calculations used separate isovector and isoscalar vector-meson form factors. He obtained excellent agreement for the proton data, but only a reasonable description of the neutron data. Kim \etal\cite{goek} used an SU(3) Nambu-Jona-Lasinio Lagrangian, an effective theory that incorporates spontaneous chiral symmetry breaking. This procedure is comparable to the inclusion of vector mesons into the Skyrme model, but it involves many fewer free parameters (which are fitted to the masses and decay constants of pions and kaons). The calculations are limited to \Q $\le$ 1 \GeV\ because the model is restricted to Goldstone bosons and because higher-order terms, such as recoil corrections, are neglected. A constituent quark mass of 420 MeV  provided a reasonable description of the EMFF data (Figure \ref{Chiral}).

In the asymptotically free limit, QCD can be solved perturbatively, providing predictions for the EMFF behavior at large \Q-values. Brodsky \& Farrar\cite{brod1} derived a scaling law for the Pauli and Dirac form factors based on a dimensional analysis, that entailed counting propagators and the number of scattered constituents:

\begin{equation} 
F_1 \propto (Q^2)^{-2}, ~~~ F_2 \propto (Q^2)^{-3}, ~~~F_2/F_1 \propto Q^{-2}
\end{equation}

Brodsy and Lepage\cite{brod2} later reached the same asymptotic behavior based on a more detailed theory that assumed factorization and hadron helicity conservation. The recent polarization transfer data clearly do not follow this pQCD prediction (which the Rosenbluth data unfortunately do). Miller\cite{mill2} was the first to observe that imposing Poincar\'e invariance removes the pQCD condition that the transverse momentum must be zero, and introduces a quark OAM component in the wavefunction of the proton, thus violating hadron helicity conservation. His model predicts a $1/Q$ behaviour for the ratio of the Dirac and Pauli form factors at intermediate \Q-values, in excellent agreement with the polarization transfer data for $Q^2 \ge 3$ \GeV. Iachello\cite{iach2} and others have pointed out that this $1/Q$ behaviour is accidental and only valid in an intermediate \Q-region.
Ralston\cite{rals} has generalized this issue to conclude that the \Q-behavior of the Jefferson Lab data signals substantial quark OAM in the proton. Recently, Brodsky \etal\cite{brod3} and Belitsky \etal\cite{beli} have independently revisited the pQCD domain. Belitsky \etal\ derive the following large \Q-behavior:

\begin{equation} 
\frac{F_2}{F_1} \propto \frac{\ln^2{Q^2/ \Lambda ^2}}{Q^2},
\end{equation}

\begin{figure}[h!]
\epsfxsize =12 cm		
\centerline{\epsfbox{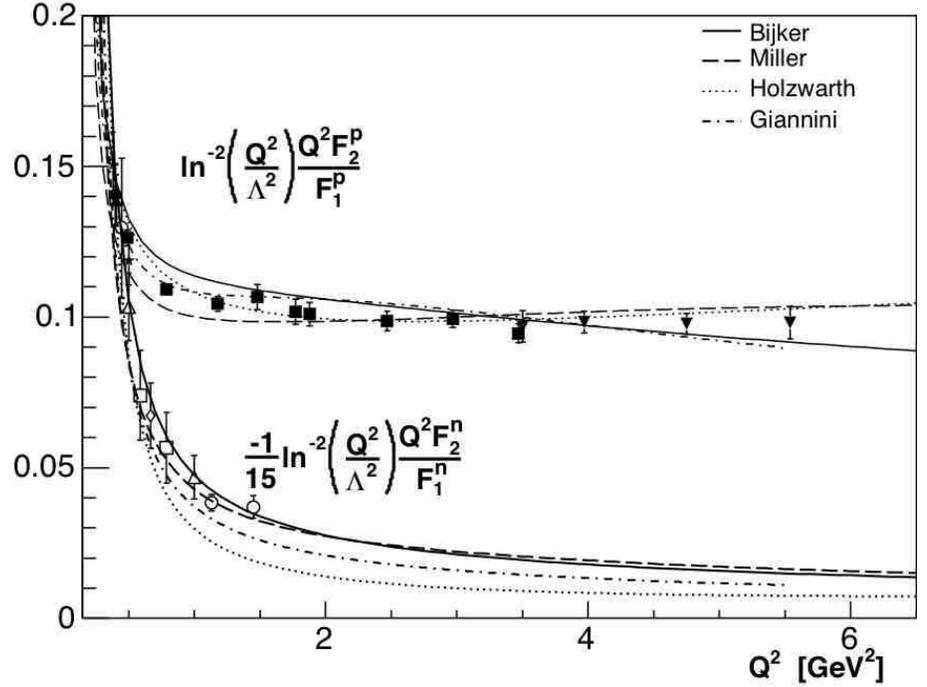}}
\caption{The ratio $(Q^2   F_2 / F_1)/ \ln^2{(Q^2/ \Lambda ^2)}$ as a function of \Q\ for the polarization-transfer data and the calculations of References \cite{iach3,mill1,holz,gian}.  The same ratio, scaled by a factor -1/15, is shown for the neutron with open symbols. For $\Lambda$ a value of 300 MeV has been used.}
\label{F2F1}
\end{figure}

\setlength{\parindent}{0em}
where $\Lambda$ is a soft scale related to the size of the nucleon. Even though the Jefferson Lab data follow this behavior (Figure \ref{F2F1}), Belitsky \etal\ warn that this could very well be precocious, since pQCD is not expected to be valid at such low \Q-values. Brodsky \etal \cite{brod3} argue that a nonzero OAM wave function should contribute to both $F_1$ and $F_2$ and that thus $Q^2 F_2 / F_1$ should still be asymptotically constant.

Once enough data have been collected on generalized parton distributions, it will become possible to construct a three-dimensional picture of the nucleons, with the three dimensions being the two transverse spatial coordinates and the longitudinal momentum. Miller\cite{mill3} has further investigated the information that can be extracted from form-factor data by themselves. His colorful images of the proton should be interpreted as three-dimensional pictures of the proton as a function of the momentum of the quark, probed by the virtual photon, and for different orientations of the spin of that quark relative to that of the proton. Ji\cite{ji} has derived similar images from generalized parton distributions using Wigner correlation functions for the quark and gluon distributions.

However, all theories described until now are at least to 
some extent effective (or parametrizations). They use models constructed to focus on certain selected aspects of QCD. Only lattice gauge theory can provide a truly ab initio calculation, but accurate lattice QCD 
results for the EMFFs are still several years away. One of the most advanced lattice calculations of EMFFs has been performed by the QCDSF collaboration\cite{gock}. The technical state of the art limits these calculations to the quenched approximation (in which sea-quark contributions are neglected), to a box size of 1.6 fm and to a pion mass of 650 MeV. Ashley \etal \cite{ashl} have extrapolated the results of these calculations to the chiral limit, using chiral coefficients appropriate to full QCD. The agreement with the data (Figure \ref{Chiral}) is poorer than that of any of the other calculations, a clear indication of the technology developments required before lattice QCD calculations can provide a stringent test of experimental EMFF data.

\section{Generalized Polarizabilities of the Nucleon}

\subsection{Introduction}

The electric and magnetic polarizabilities of an object describe how its internal structure
responds to external electric and magnetic fields.  
In the weak field limit, an external electric field 
{\bf E} induces in a finite system an electric dipole moment {\bf p}
proportional to the applied field:
\begin{eqnarray}
{\bf p} &=& \alpha_E {\bf E}
\end{eqnarray}
(Heaviside-Lorentz units).
This proportionality defines the electric polarizability $\alpha_E$.
The induced dipole moment ${\bf p}$ is measurable by the effects of the
long range dipole electric field it produces.  Similarly, a weak external
magnetic field {\bf H} induces a magnetic dipole moment
\begin{equation}
\Delta\vec{\mu} = \beta_M {\bf H},
\end{equation}
which defines the magnetic polarizability $\beta_M$.
The induced moment $\Delta\vec{\mu}$ is a change in the static moment
${\bf \mu}_0$ of a non spin zero system.

It is instructive
to contrast a few basic examples of the polarizabilities. For the hydrogen atom, the
electric polarizability is of the same order of magnitude as the
hydrogen atom volume, whereas the proton's polarizability is
roughly $10^{-3}$ of its volume.  Thus the proton is much
stiffer than the hydrogen atom.
In the Schr\"{o}dinger
equation for the hydrogen atom, the electric polarizability is
\cite{Holstein:1999uu}:
\begin{eqnarray}
\alpha_E^{\rm H\ atom} &=& {27 \over 8 \pi} 
\left[ {4\over 3} \pi a_0^3 \right],
\end{eqnarray}
where $a_0 = 1 /(m_e\alpha_{QED})$ is the Bohr radius.
For a particle of mass $m$ and charge $e$ 
bound by a simple harmonic oscillator of length
constant $b$, the electric polarizability is
\begin{eqnarray}
\alpha_E^{H.O.} &=& \alpha_{QED} \left[b \over \lambda_C \right] b^3,
\label{eq:alphaHO}
\end{eqnarray}
where $\lambda_C = 1 / m$ is the Compton wavelength of the particle.
The Hydrogen atom result is compatible with a non-relativistic 
harmonic oscillator model with $b\approx a_0$ and therefore 
$b/\lambda_C\gg 1$.
On the other hand, a harmonic oscillator model of the small value
of the proton polarizability requires $b/\lambda_C \approx 1$
for the quarks or pions of the proton substructure.
Thus the order of magnitude
of the nucleon polarizabilities is evidence for the
explicitly relativistic structure of the nucleon.  This illustrates how
 the polarizabilities and their $Q^2$ dependent
generalizations reveal details of the nucleon dynamics that go 
beyond the charge and magnetization distributions of the form factors.

\subsection{Proton Polarizabilities}

The low-energy limit of the Compton amplitude is determined by the
nucleon's charge and magnetic moment.  As the energy of the probe
increases, the effect of the excitation spectrum of the nucleon
can be summarized by a set of electromagnetic polarizabilities.
Today there is an extensive experimental program to measure the 
proton polarizabilities in real and virtual Compton scattering and
the neutron polarizabilities in elastic and quasi-elastic
Compton scattering on the deuteron.

The scattering amplitude for elastic real photon scattering from
a nucleon can be described by six complex amplitudes, each of which
multiplies a linearly independent 
algebraic structure of the polarization and kinematic variables 
\cite{Bardeen:aw}.  Below pion production threshold these amplitudes are real,
to lowest order in $\alpha_{QED}=e^2/(4\pi)$.

To illustrate the low energy structure of Compton scattering,
consider first the forward Compton amplitude on a nucleon of charge
$\lambda e$
 and anomalous magnetic moment $\kappa$.
The scattering amplitude has the following form 
\cite{Petrunkin:1981me,Babusci:1998ww}:
\begin{eqnarray}
{1 \over 8\pi M_N }\hat{T}(0^\circ) &=& 
\vec{ \epsilon}^{\,\prime *} \cdot
  \vec{\epsilon}f_1(\omega^2) + i \omega  
  \vec{\sigma}\cdot( \vec{ \epsilon}^{\,\prime *} \times  \vec{\epsilon} )
  f_2(\omega^2). \\
  &=&
\vec{ \epsilon}^{\,\prime *} \cdot
  \vec{\epsilon}\left[-\lambda^2{\alpha_{QED}\over M_N}
  +\omega^2(\alpha_E+\beta_M)
  +{\mathcal O}(\omega^4)\right] \nonumber\\ 
 & &
  - i \omega  \vec{\sigma}\cdot( \vec{ \epsilon}^{\,\prime *} \times  \vec{\epsilon} )
  \left[{\kappa^2\alpha_{QED}\over 2 M_N^2} - \gamma_0\omega^2 +{\mathcal O}(\omega^4)\right].\label{eq:RCSamp}
\end{eqnarray}
Here $\omega$ is the laboratory photon energy, $\vec{\epsilon}$
and $\vec{ \epsilon}^{\,\prime }$ the initial and final photon polarization
vectors, and $\vec{\sigma}$ the  nucleon spin operator.
The forward spin polarizability $\gamma_0$ will be  discussed later.

To order $\omega^2$, the spin-averaged Compton cross section  is given by
\cite{Petrunkin:1981me,Babusci:1998ww}:
\begin{eqnarray}
{d\sigma(\gamma,\gamma)\over d\Omega_{\gamma\gamma}^{\rm lab} } &=&
{1\over 4} \sum_{f,i}\left|
\left< f \right| {1\over 8\pi M_N}{\omega'\over\omega}\hat{T} \left| i \right>
\right|^2 \nonumber \\
&=&
{d\sigma^B(\gamma,\gamma)\over d\Omega_{\gamma\gamma} }
-{\lambda^2 \alpha_{QED} \over M_N} \left[{\omega'\over \omega}\right]^2
\nonumber \\
& & \times \left[ {\alpha_E+\beta_M\over 2}(1+\cos\theta)^2+
                  {\alpha_E-\beta_M\over 2}(1-\cos\theta)^2\right].
 \label{eq:RCS}
\end{eqnarray}
${d\sigma^B/ d\Omega_{\gamma\gamma}}$ is the exact (Born)
cross section for a nucleon, given  by Powell \cite{ref:Powell}.  It differs from
the Klein-Nishina formula\cite{ref:KleinN}{}
  in the inclusion of the anomalous magnetic
moment, $\kappa$.
The linear dependence of the cross section on the polarizabilities
$\alpha_E$ and $\beta_M$ at order $\omega^2$ results from the interference 
of the
${\mathcal O}(\omega^2)$ amplitude with the order unity Thomson amplitude.
For Compton scattering from a free neutron, since $\lambda=0$, this
interference is absent and the polarizabilities
enter the cross section at order $\omega^4$.

Each of the six independent terms in the Compton amplitude can be
constrained by dispersion relations (DR).  For the forward Compton amplitude,
these relations take a particularly simple form, since the 
optical theorem connects the imaginary part of the forward scattering
amplitude to the photo-absorption cross section.
For $f_2$, the dispersion relation yields the 
Gerasimov-Drell-Hearn (GDH) sum rule
\cite{Drell:1966jv,Gerasimov:1966et}:
\begin{eqnarray}
{\alpha_{QED}\over M^2} \kappa^2 &=& {1\over 2 \pi^2}
\int_{\rm Th}^\infty \left[\sigma_{3/2}(\omega)-\sigma_{1/2}(\omega)
\right]
{d\omega\over \omega.}  \label{eq:GDH}
\end{eqnarray}
For $f_1$, the subtracted dispersion relation yields the Baldin
sum rule \cite{ref:Baldin}
\begin{eqnarray}
\alpha_E + \beta_M = {1\over 2 \pi^2}
     \int_{\rm Th}^\infty {\sigma_\gamma(\omega)\over \omega^2} d\omega 
    &=& [14.20 \pm 0.50]\cdot 10^{-4}\,{\rm fm}^3\ \ (1970),
\label{eq:Baldin}\\
    &=& [13.69 \pm 0.14]\cdot 10^{-4}\,{\rm fm}^3\ \ (1998). \label{eq:BB}
\end{eqnarray}
The $1/ \omega^2$ behavior of the integrand of Equation~\ref{eq:Baldin}
 gives a rapid convergence of the integral.  Already in 1970, Damashek \&
Gilman \cite{ref:Damashek}{} were able to obtain the value quoted in 
Equation~\ref{eq:Baldin}.  The
intervening 30 years of experiments have resulted in the improved analysis of
Babusci et al.{}
\cite{Babusci:1997ij} in Equation~\ref{eq:BB}.
This convergence is to be contrasted with the much more
difficult problem of the experimental 
evaluation of the GDH sum rule (Equation~\ref{eq:GDH}), since the 
experimental integration saturates only if a model is used for the (unmeasured)
high energy behavior of the integrand.
In contrast with Equation~\ref{eq:Baldin}, the dispersion relation for $\alpha_E-
\beta_M$ is not  convergent.  Thus, this combination
of polarizabilities must be determined directly from low energy Compton
scattering.
%
In an effective theory, the lowest order interaction of the external
electromagnetic field with the internal structure of the proton
(beyond the charge and magnetic moment)
is entirely described by the static electric and magnetic polarizabilities:
\begin{eqnarray}
{\mathcal H}_{\rm Int}^{(2)} &=& -{1\over 2}
\left[4\pi \alpha_E {\bf E}^2 + 4\pi \beta_M {\bf H}^2\right].
\label{eq:static}
\end{eqnarray}
L'vov  showed that the  ${\mathcal O}(\omega^2)$ 
terms in the scattering amplitude are the same
polarizabilities \cite{Lvov:1993fp}.

Starting with Gol'danski \cite{Goldanski:1960}, several generations of 
tagged and untagged bremsstrahlung experiments have
tackled the measurement of low energy Compton scattering from the proton
\cite{Baranov:1975ju,MacGibbon:1995in,Hallin:1993ft,Zieger:1992jq,Federspiel:1991yd};
see \cite{ref:Nathan} for a historical review.
The bremsstrahlung photon beam is obtained by passing the primary electron
beam through a high $Z$ radiator (such as
Cu or W), a few percent of a radiation length in thickness.  
The electron beam is deflected by a magnet so that only the
photons reach the hydrogen production target.  
In the tagged experiments,  the bremsstrahlung electrons are detected
in coincidence with the scattered photon, thereby ``tagging'' the incident
photon energy of the Compton event.  In the untagged experiments,
the incident photon energy is determined event-by-event from the
constrained kinematics of the elastic scattering process.

The most recent and precise Compton results are from the MAMI accelerator
 \cite{OlmosdeLeon:zn}.
 In order to extract   the 
polarizabilities from Compton scattering data at and above pion threshold,
the higher order terms in the full scattering amplitude are constrained
by dispersion relations, derived  in particular
 by  L'vov, and collaborators
\cite{ref:Lvov,ref:Schumacher}.
The dispersion relations are evaluated by pion photo-production multipole
analysis along with a Regge theory extrapolation for the asymptotic
part.  Drechsel \etal{} \cite{Drechsel:1999rf,Drechsel:2002ar}{}
also discuss subtracted dispersion
relations for two of the six amplitudes.

A global analysis of the Compton data yielded the following values
for the polarizabilities \cite{OlmosdeLeon:zn}:
\begin{eqnarray}
\alpha_E &=& [12.1\pm 0.3 (stat.) \mp 0.4 (syst.) \pm 0.3 (model)] \cdot 10^{-4}\ {\rm fm}^3 \nonumber\\
\beta_M  &=& \  [1.6\pm 0.4 (stat.) \pm 0.4 (syst.) \pm 0.4(model)] \cdot 10^{-4}\ {\rm fm}^3. 
\label{eq:Olmos}
\end{eqnarray}
The systematic errors are anti-correlated from the Baldin sum rule constraint
(Equation~\ref{eq:BB}).  The model dependent errors are  estimates of the uncertainties
from the dispersion relations.

The ${\mathcal O}(\omega^3)$ spin-dependent and 
${\mathcal O}(\omega^4)$ terms in the Compton scattering amplitude can
also be connected to polarizabilities, as discussed in detail by
Babusci \etal{}
\cite{Babusci:1998ww}.  In
the presence of time- or spacedependent electromagnetic fields, the internal
structure of the nucleon will polarize in response to the form of the external
field. The most general effective interaction (to order $\omega^4$) between
the external field and the internal nucleon structure is given by
\cite{Babusci:1998ww}:
\begin{eqnarray}
{\mathcal H}_{\rm Int} &=& {\mathcal H}_{\rm Int}^{(2)}+ 
{\mathcal H}_{\rm Int}^{(3)} + {\mathcal H}_{\rm Int}^{(4)}. 
 \label{eq:H234}\\
{\mathcal H}_{\rm Int}^{(3)} &=& -{4 \pi} 
\left[{\gamma_{E1}\over 2}  {\vec{\sigma}}\cdot{\bf E}\times{\bf \dot{E}} +
      {\gamma_{M1}\over 2}  {\vec{\sigma}}\cdot{\bf H}\times{\bf \dot{H}} -
       \gamma_{E2} {{\sigma_i}E_{ij}}H_j +
       \gamma_{M2} {\sigma_i}{H_{ij}}E_j
\right], \nonumber \\ 
{\mathcal H}_{\rm Int}^{(4)} &=& -{4 \pi\over 2} 
\left[\alpha_{E\nu} {\bf \dot{E}}^2 +
       \beta_{M\nu} {\bf \dot{H}}^2 +
     {1\over 6}( \alpha_{E2} E_{ij}^2+
                 \beta_{M2} {H_{ij}}^2 )
\right],  \label{eq:SpinPol}
\end{eqnarray}
where $E_{ij} = (\nabla_iE_j + \nabla_jE_i)/2$ and similarly for
$H_{ij}$.  Because of the extra space and time derivatives in
Equation~\ref{eq:SpinPol}, the polarizabilities defined by 
${\mathcal H}_{\rm Int}^{(3)} $ and
${\mathcal H}_{\rm Int}^{(4)} $ enter the spin-dependent
and spin-independent Compton scattering amplitudes to order
$\omega^3$ and $\omega^4$, respectively.
The polarizabilities $\gamma_{E2}$ and $\gamma_{M2}$ measure the
spin dependent quadrupole strength in the nucleon spectrum.
DR estimates and theoretical calculations of these
higher order polarizabilities
are presented in References
\cite{Babusci:1998ww} and 
\cite{Holstein:1999uu}.

Of particular experimental 
interest are the forward $\gamma_0$ and backward $\gamma_\pi$
spin polarizabilities:
\begin{eqnarray}
\gamma_0 &=&  - \gamma_{E1} - \gamma_{M1} - \gamma_{E2}-\gamma_{M2} \\
 &=& - {1\over 4\pi^2}\int {d\omega\over \omega^3}
\left[\sigma_{3/2}(\omega)-\sigma_{1/2}(\omega)\right] \ \ 
\cite{Gell-Mann:db,Gell-Mann:1954kc}  \label{eq:gamma0}
\\
\gamma_\pi &=&  - \gamma_{E1} + \gamma_{M1} + \gamma_{E2}-\gamma_{M2}
\end{eqnarray}
The forward spin polarizability $\gamma_0$ 
is the ${\mathcal O}(\omega^3)$ term
in the forward scattering amplitude of Equation~\ref{eq:RCSamp}.
Similarly, $\gamma_\pi$ is the  ${\mathcal O}(\omega^3)$ spin-dependent
term in the amplitude for Compton scattering in the backward direction
($\theta_{\gamma\gamma}=\pi$).

Sandorfi \etal{} extracted the forward spin polarizability 
 from a dispersion
analysis of photoproduction multipoles \cite{Sandorfi:ku}.
The MAMI\cite{Ahrens:2001qt} and ELSA\cite{Dutz:mm}
GDH experiments, measured the following contributions 
(units $10^{-4}$ fm$^4$), respectively,
$\gamma_0 = -1.87\pm 0.08\pm 0.10$ for 
0.2 GeV $\le \omega\le 0.8$ GeV and
$\gamma_0 = -0.027\pm 0.002\pm  0.001$ for 0.8 GeV $\le \omega \le 
1.82$ GeV.  From this, along with the multipole estimate 
$\gamma_0 = +0.90$ for $\omega < 0.2$ GeV
\cite{Drechsel:2002ar,Tiator:2002zj}{} one obtains:
\begin{eqnarray}
\gamma_0 &=& [-1.02\pm 0.08(stat)\pm 0.10(syst)]\cdot 10^{-4} {\rm fm}^4
\end{eqnarray}

There are several measurements of the backward spin polarizability 
$\gamma_\pi$.
 The overall error envelopes in the analysis of the low energy
Compton data (Equation~\ref{eq:Olmos}) increase slightly if  $\gamma_\pi$
is included as a fitting parameter, with the result \cite{OlmosdeLeon:zn}:
\begin{eqnarray}
\gamma_\pi &=& [-36.1\pm 2.1 (stat.) \mp 0.4 (syst.) \pm 0.8 (model)] \cdot 10^{-4}\ {\rm fm}^4.
\label{eq:Olmos_dpi}
\end{eqnarray}
The LEGS group obtained the result \cite{Tonnison:1998mi,Blanpied:2001ae}
\begin{eqnarray}
\gamma_\pi &=& [-27.1\pm 2.3 (stat.+syst) \pm 2.2 (model)] \cdot 10^{-4}\ {\rm fm}^4
\label{eq:LEGS_dpi}
\end{eqnarray}from back-scattering $\gamma p \rightarrow \gamma p$ and
$\vec{\gamma} p \rightarrow \gamma p$ 
data in the $\Delta$-resonance region.
However, new Compton cross section measurements in the $\Delta$-resonance
confirm the larger value \cite{Galler:2001ht,Wolf:2001ha}
\begin{eqnarray}
\gamma_\pi &=& [-37.1\pm 0.6 (stat.+syst) \pm 3.0 (model)] \cdot 10^{-4}\ {\rm fm}^4.
\label{eq:Wolf_dpi}
\end{eqnarray}
This result was again confirmed with a second  apparatus at MAMI
\cite{Camen:2001st}.
When one compares the results of Equations~\ref{eq:Olmos_dpi} and \ref{eq:Wolf_dpi}, 
the high energy data yield an improved statistical precision at the 
expense of a greater
model uncertainty.  
The model uncertainties include distinct analysis with 
the MAID\cite{Drechsel:1998hk} or SAID \cite{Arndt:2002xv}
 multipoles,  and variations within
each multipole parameterization.

The $\gamma\gamma\pi^0$ triangle anomaly  dominates the backward
spin polarizability via the $\pi^0$ $t$-channel
exchange\cite{L'vov:1999ez}:
\begin{equation}
\gamma_\pi^{(\pi^0)} = -45.0 \pm 1.6 \cdot 10^{-4}\ {\rm fm}^4.
\end{equation}
This contribution is absent from $\gamma_0$.
The larger experimental value of $|\gamma_\pi^p|$
is consistent with DR calculations
\cite{Drechsel:1997xx,L'vov:1999ez}.
In addition to the $\pi^0$ anomaly, L'vov and Nathan  \cite{L'vov:1999ez}
compute contributions of $+7.3$, $-0.3$, and $-1.6$ (units
of $10^{-4}$ fm$^4$)  from $\pi N$ $s$-channel,
$\pi\pi N$ $s$-channel, and $\eta,\eta^\prime$ $t$-channel
contributions, respectively.  Their total DR prediction is 
$\gamma_\pi = (-39.5\pm 2.4)\cdot 10^4$ fm$^{4}$.  This calculation
also illustrates that in contrast with $\alpha_E-\beta_M$, for
which Regge phenomenology suggests the dispersion relations do not
converge, or $\alpha_E+\beta_M$, for which the
$\pi\pi N$ channel contributes 15\% to the Baldin sum rule, the
dispersion relations for the higher order polarizabilities converge
rapidly.

\subsection{Theoretical Perspective}
\label{sec:RCStheory}

In Heavy Baryon Chiral Perturbation Theory (HBChPT) to ${\mathcal O}(p^3)$,
the polarizabilities have a very simple form 
\cite{Holstein:1999uu,Bernard:1992}:
\begin{eqnarray}
 \alpha_E^p=\alpha_E^n &=& 5 \alpha_{QED} g_A^2 / [ 96 \pi m_\pi F_\pi^2]
=12.5 \cdot 10^{-4} {\rm fm}^3 \nonumber \\
\beta_M^p = \beta_M^n &=& \alpha_E/10 \\
\gamma_0^p=\gamma_0^n &=& {8\over 10} {1\over \pi m_\pi} \alpha_E^p 
            = 4.52\cdot 10^{-4} {\rm fm}^4                    \\
\gamma_\pi^p  &=& -\gamma_0^p\left[{12\over g_A} - 1\right] 
 = - 38.3\cdot 10^{-4} {\rm fm}^4, \\
 \gamma_\pi^n &=& \gamma_0^p\left[{12\over g_A} + 1\right]  
 = + 47.3\cdot 10^{-4} {\rm fm}^4, 
\end{eqnarray}
where $g_A \approx 1.266$ is the nucleon axial coupling constant
and $F_\pi\approx 92.4$ MeV is the pion decay constant.
The HBChPT results are in remarkable agreement with the proton data.
Even for $\gamma_0^p$, if the $\pi^0$ anomaly term 
$12 \gamma_0 /g_A$ in $\gamma_\pi$
is considered to set the scale for each of the four terms in
Equation~\ref{eq:gamma0}, the disagreement between theory and experiment
is small.

Bernard \etal{} \cite{Bernard:1993bg,Bernard:1993ry} 
obtained the following values for  $\alpha_E$ and $\beta_M$ of the nucleon to
${\mathcal O}(p^4)$  in HBChPT:
\begin{eqnarray}
\alpha_E^p = (10.5 \pm 2.0) \cdot 10^{-4}\, {\rm fm}^3&\phantom{1234}&
\beta_M^p  = ( 3.5 \pm 3.6) \cdot 10^{-4}\, {\rm fm}^3  \nonumber \\
\alpha_E^n = (13.4 \pm 1.5) \cdot 10^{-4}\, {\rm fm}^3& & 
\beta_M^n  = ( 7.8 \pm 3.6) \cdot 10^{-4}\, {\rm fm}^3. \label{eq:HBChPTO4}
\end{eqnarray}
The theoretical error bars come from the phenomenological constants
in the effective Lagrangian, especially the terms governing the 
$\Delta$-Resonance contribution.  Hemmert \etal{} \cite{Hemmert:1996rw}{}
introduced
the $\Delta$ as an explicit degree of freedom in the 
 the small-scale expansion (SSE) 
in which the mass difference $M_\Delta-M_N$ is treated as an expansion
parameter to ${\mathcal O}(p^3)$ along with $m_\pi$ and the momentum $p$.  
They obtain a large ($4-7\cdot 10^{-4}$ fm$^3$) additional
positive contribution to $\alpha_E$ from $\Delta \pi$ loops and to
$\beta_M$ from the $\Delta$-pole (see below, Equation \ref{eq:SSE}).  
These effects are expected to
be canceled in higher orders, as suggested in Reference \cite{Bernard:1993bg}.
For the spin polarizabilities $\gamma_{E1,M1,E2,M2}$, 
the ${\mathcal O}(p^4)$ ChPT calculations
do not introduce new free parameters 
\cite{Ji:1999sv,Kumar:2000pv,Gellas:2000mx,Birse:2000ve,Gellas:yv}.
 However, the 
values for $\gamma_0$ and $\gamma_\pi$ change by approximately 
the magnitude of $\gamma_0$, relative to the ${\mathcal O}(p^3)$ results.

These theoretical results leave somewhat in question the
convergence of the chiral expansions for the polarizabilities.
However, combined with the data, they confirm basic expectations.
The magnetic polarizability has a strong cancellation between paramagnetism
(from the constituent quark spin flip, or equivalently $N\rightarrow N^*$
terms) and diamagnetism driven by the pion cloud. For the
electric polarizability, both degrees of freedom  
contribute with the same sign.  It remains for the study of 
generalized polarizabilities to see if these contributions have
a different spatial structure.

Instead of attempting to predict the polarizabilities, 
Beane \etal{}\cite{Beane:2002wn}{} 
used the
electric and magnetic polarizabilities as the only two
free parameters in a 
${\mathcal O}(p^4)$ chiral perturbation expansion of the Compton amplitude
\cite{McGovern:2001rx}.  After refitting the data below $200$ MeV, they obtain
\begin{eqnarray}
\left\{\alpha_E^p \atop \beta_M^p \right\} &=& 
\left\{12.1 \pm 1.1\pm 0.5  \atop 3.4 \pm 1.1 \pm 0.1 \right\} 
\cdot 10^{-4}\, {\rm fm}^3
\ \ \hbox{(Chiral Fit),} \label{eq:ChFit}
\end{eqnarray}
consistent with the DR analysis of the same data, but with
slightly larger uncertainties.

Magnetic polarizabilities for hadrons were calculated in Lattice QCD 
by Zhou \etal{} \cite{Zhou:2002km}, 
for values of $0.3 \le m_\pi^2 \le 1  $ GeV$^2$.  They added a 
static magnetic field to the lattice, and extracted the polarizability
from the quadratic dependence of the ground state mass on the external field
(Equation~\ref{eq:static}).
Substantially greater computational resources are needed to obtain results
closer to the physical pion mass, but at the scale $m_\pi^2=0.3$ the
results are $\beta_M^p = 0\pm 1$ and $\beta_M^n = 11\pm 1$.
Christensen \etal{} \cite{Christensen:2002wh}{}
 used the same technique to obtain electric
polarizabilities of neutral hadrons.  The result for $\alpha_E^n$ shows
considerably larger numerical uncertainty than the $\beta_M^n$ result.

\subsection{Neutron Polarizabilities}

Wissmann \etal{} reviewed the experimental techniques for
measuring the neutron polarizabilities \cite{Wissmann:1998ta}.
Two recent analysis of the total photo-absorption data on the deuteron
and $(\gamma,\pi)$ threshold multipoles give the following values
for the Baldin sum rule of the neutron polarizabilities:
\begin{eqnarray}
\alpha_E^n + \beta_M^n&=& 14.40 \pm 0.66 \cdot 10^{-4}\, {\rm fm}^3 
\ \ \cite{Babusci:1997ij}\\
\alpha_E^n + \beta_M^n&=& 15.2 \pm 0.5 \cdot 10^{-4}\, {\rm fm}^3
\ \ \cite{Levchuk:1999zy}.
\end{eqnarray}
Schmiedmayer \etal{} \cite{Schmiedmayer:bm} and
 Koester \etal{} \cite{Koester:nx}{} extracted
the neutron electric polarizability  from
the energy dependence of low energy neutron-nucleus scattering.
Their results are:
\begin{eqnarray}
\alpha_n &=& \left[12.6 \pm 1.5\pm 2.0\right] \cdot 10^{-4}\, {\rm fm}^3
\ \ \cite{Schmiedmayer:bm} \ \ \\
\alpha_n &=& \left[0.6 \pm 5 \right] \cdot 10^{-4}\, {\rm fm}^3
\ \ \cite{Koester:nx}. \ \ 
\end{eqnarray}
Koester et al. \cite{Koester:nx} and Enik \etal{}
\cite{Enik:cw}
suggest the uncertainty in both measurements 
should  be 
 $\pm 5\cdot 10^{-4} $ fm$^3$. 
The
neutron polarizabilities were also  extracted from
$D(\gamma, \gamma) D$ and $ D (\gamma, \gamma  n) p$
 measurements at MAMI-A \cite{Rose:eu,ref:Rose}
the Saskatoon Accelerator Laboratory (SAL)
\cite{Hornidge:1999xs,Kolb:2000ix},
Lund MAX-Lab \cite{Lundin:2002jy} and MAMI \cite{Kossert:2002jc}.

The most precise quasi-elastic $D(\gamma,n)p$ data were obtained
at MAMI, with the result\cite{Kossert:2002jc,Kossert:2002ws}
\begin{eqnarray}
\alpha_E^n -\beta_M^n &=& 9.8 \pm 3.6 ({stat}){+2.1 \atop -1.1}({ syst})
\pm 2.2 ({model}) \cdot 10^{-4}\, {\rm fm}^3. 
\end{eqnarray}
The Mainz experiment also extracted the polarizabilities of the bound
proton via the $D (\gamma, \gamma  p) n$ reaction, with
the result \cite{Wissmann:vi}:
\begin{eqnarray}
[\alpha_E^p - \beta_M^p]^{\rm bound} &=& 9.1 \pm 1.7 ({ stat + sys})\pm
1.2 ({ model}) \cdot 10^{-4}\, {\rm fm}^3. 
\end{eqnarray}
Although they did not extract a value for $\gamma_\pi$ for
a bound proton, the authors note that their analysis is consistent
with the free value of $\gamma_\pi = -37.6$ and inconsistent 
with  $\gamma_\pi = -27.1$ (units of $10^{-4}$ fm$^4$).
The agreement with the free proton values for $\alpha_E - \beta_M$ and
$\gamma_\pi$ confirms
the basic validity of the theoretical framework for the
extraction of nucleon polarizabilities from 
 quasi-free Compton scattering on the deuteron
\cite{Levchuk:1994ij}.

Hornidge \etal{} \cite{Hornidge:1999xs} and 
Lundin \etal{} \cite{Lundin:2002jy} measured the coherent Compton scattering
on the deuteron:
$\gamma D \rightarrow \gamma D$.  The data from SAL are at the
highest energy (94 MeV) and therefore have the greatest sensitivity to
the polarizabilities, but they are  integrated over a 20 MeV energy
bin, compared to the 10 MeV bins of the Lund data.
From a global analysis
using the  $NN$-potential model formalism
of Levchuk \& L'vov \cite{Levchuk:1999zy},
Lundin \etal{} extract the isoscalar polarizabilities
\begin{eqnarray}
\alpha_E^N + \beta_M^N &=& 16.7 \pm 1.6 \cdot 10^{-4}\, {\rm fm}^3\nonumber \\
\alpha_E^N - \beta_M^N &=& 4.8 \pm 2.0  \cdot 10^{-4}\, {\rm fm}^3
\end{eqnarray}
  The theoretical formalism for deuteron Compton scattering
is also discussed in Reference \cite{Karakowski:1999eb}.
As already described for the proton Compton case, Beane \etal{} reanalyzed
the data in a chiral expansion of the scattering amplitude, and obtained
\cite{Beane:2002wn}
\begin{eqnarray}
\alpha_E^N &=& 9.0 \pm 1.5 {+3.6 \atop -0.8} \cdot 10^{-4}\, {\rm fm}^3
 \nonumber \\
\beta_M^N &=& 1.7 \pm 1.5 {+1.4 \atop -0.6} \cdot 10^{-4}\, {\rm fm}^3.
\end{eqnarray}
The error bars result from both model uncertainties and inconsistencies
between the data sets, with the chiral analysis more consistent with the
Lund data alone.

Measuring neutron polarizabilities to the same precision as the
proton remains a formidable challenge.

\subsection{Virtual Compton Scattering}

\begin{figure}[h!]
\epsfxsize=\textwidth
\centerline{\epsfbox{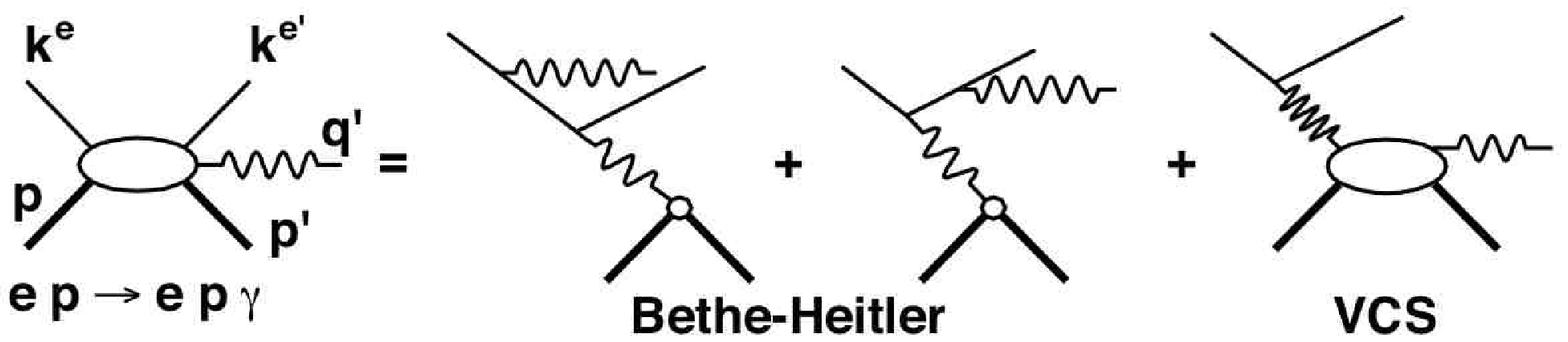}}
\vskip 1em
\epsfxsize=\textwidth
\centerline{\epsfbox{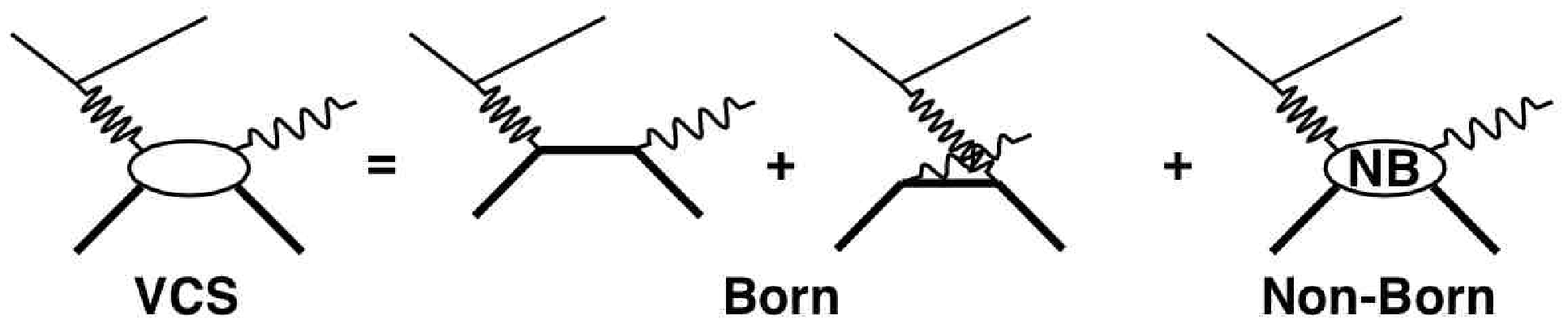}}
\caption{Kinematics and scattering amplitude 
for the $e p \rightarrow e p \gamma$ reaction.
The incident and scattered electron 4-momentum vectors are
$k^e$ and $k^{e\,\prime}$, respectively.  The initial and
final proton 4-momentum vectors are 
$p$ and $p'$, respectively.  
The final photon and the VCS virtual photon 4-momentum vectors are $q'$
and $q_\mu=(k^e-k^{e\prime})$, respectively.
The Mandelstam invariants are $s = (q+p)^2$, 
$t=(q - q^\prime)^2$, and $u=(p-q^\prime)^2$.}
\label{fig:VCSkin}
\end{figure}

Virtual Compton Scattering (VCS) can be measured in the
$e p \rightarrow e p \gamma$ reaction. 
  In this case, the Compton amplitude
interferes with the Bethe-Heitler (electron radiation) amplitude,
as illustrated in Figure~\ref{fig:VCSkin}, which also defines the kinematic variables. 
Guichon \etal{}  constructed a gauge invariant 
separation of   the $e p \rightarrow e p \gamma$ amplitude
into the Bethe-Heitler (BH), Born  and Non-Born (NB) terms
\cite{Guichon:1995pu}.   The BH and Born terms are the 
amplitudes for electron and proton bremsstrahlung, including only the
on-shell proton form factors $F_{1,2}$. 
They  then expanded the  amplitude in
 powers of $q'$,
the final photon energy in the proton-photon center-of-mass frame. 
The leading term in the expansion
of the BH and Born amplitudes is ${\mathcal O}(1/q')$,
 arising from the electron and proton propagators,
respectively.  The leading term in the Non-Born amplitude is 
$ {\mathcal O}(q')$.  This term is defined by 
6 generalized polarizabilities, representing the
 independent  multipoles coupling the initial virtual photon with
a final $E1$ or $M1$ photon \cite{Guichon:1995pu}.  
The  generalized polarizabilities
are functions of $\tilde{Q}^2$, the invariant momentum transfer
squared in the $q'\rightarrow 0$ limit.
The
generalized 
polarizabilities describe the spatial variation of the polarization
response of the proton, as described explicitly in Reference
 \cite{L'vov:2001fz}.

  Metz \& Drechsel applied crossing  and charge-conjugation
symmetry in the linear-$\sigma$
model, and obtained four constraints among the ten
low energy VCS multipoles
\cite{Metz:fn,Metz:1997fr}.
Later, Drechsel \etal{} showed that these constraints are general,
establishing that there are six independent generalized polarizabilities
\cite{Drechsel:1996ag,Drechsel:1997xv}.
This is a  nice application of  model building:
an approximate model incorporating chiral symmetry and exact
relativistic dynamics led to a deeper
understanding of the fundamental dynamics.

The generalized polarizabilities are labeled by
$P^{(\rho'L',\rho L)S}(Q^2)$, where
$L'$ and $L$ denote  final and initial multipolarity, respectively; 
$\rho'$ and $\rho$ indicate the polarization 
of the final and initial photon, respectively,
which may be 
Coulomb (C), magnetic (M), or electric (E); 
and $S=0,1$ for a scalar or vector
operator in nucleon spin space.  Siegert relations
connect the electric and Coulomb multipoles
\cite{Guichon:1998xv}.
The 
independent set of generalized polarizabilities that enter the
cross section to lowest order (beyond the BH+Born terms), and their
$\tilde{Q}^2\rightarrow 0$ limits, are
\cite{Guichon:1998xv}:
\begin{eqnarray}
\alpha_{QED} P^{(C1,C1)0}(\tilde{Q}^2) & \rightarrow& -\sqrt{2 /3} \,
 { \alpha_E } \nonumber\\
\alpha_{QED} P^{(M1,M1)0}(\tilde{Q}^2) & \rightarrow& -\sqrt{8 / 3} \,
\beta_M \nonumber\\
P^{(C1,C1)1}(\tilde{Q}^2)\ , \ P^{(M1,M1)1}(\tilde{Q}^2) &\rightarrow& 0
\nonumber\\
\alpha_{QED} P^{(M1,C2)1}(\tilde{Q}^2) & \rightarrow&-\left(2/3\right)^{3/2}
\gamma_{E2} \nonumber\\
\alpha_{QED} P^{(C1,M2)1}(\tilde{Q}^2)&\rightarrow&-(\sqrt{2} / 3) \,
\gamma_{M2} \label{eq:GPs}
\end{eqnarray}
In an unpolarized VCS experiment the cross section 
(to order $q^{\prime\, 0}$) has the form:
\begin{eqnarray}
d\sigma &=& d\sigma^{\rm BH+Born} + v_{LL}\left[
P_{LL} - P_{TT}/\epsilon \right] + v_{LT} P_{LT}, \label{eq:VCSsig}\\
P_{LL} &=& -\sqrt{24}M \,\GEp(\tilde{Q}^2) \,P^{(C1,C1)0}(\tilde{Q}^2)
= {4 M\over \alpha_{QED}} \GEp(\tilde{Q}^2)\alpha_E(\tilde{Q}^2) 
\label{eq:PLL}\\
P_{TT} &=& 6M(1+\tilde{\tau})  \GMp(\tilde{Q}^2)  \left[
P^{(M1,M1)1}(\tilde{Q}^2)+\sqrt{8}\tilde{\tau} 
P^{(C1,M2)1}(\tilde{Q}^2) \right]
\label{eq:PTT}\\
P_{LT} &=& \sqrt{3\over 2} { M \sqrt{1+\tilde{\tau}} }\left[
          \GEp(\tilde{Q}^2)   P^{(M1,M1)0}(\tilde{Q}^2) - 
\sqrt{6} {\GMp(\tilde{Q}^2)  }P^{(C1,C1)1}(\tilde{Q}^2)\right] 
\nonumber\\
 & =&
- {2 M \over \alpha_{QED}}  \sqrt{1+\tilde{\tau}}
    \GMp(\tilde{Q}^2) \beta_M(\tilde{Q}^2)
 - {\rm spin}\label{eq:PLT}
\end{eqnarray}
where $v_{LL}$ and $v_{TT}$ are kinematic factors defined in 
Equations 97--100 of Reference \cite{Guichon:1998xv},
$\epsilon= 1/[1+2({\bf q}^2 / Q^2)\tan\theta/2]_{\rm lab}$ is the
virtual photon polarization (in the VCS amplitude), and 
$\tilde{\tau} = \tilde{Q}^2/(4M^2)$.
The generalized polarizabilities enter 
linearly in the cross section to this order, owing to
the interference between the NB and BH+Born terms.
In the $\tilde{Q}^2\rightarrow 0 $ (RCS) limit, 
$P_{LL}\rightarrow 4 M \alpha_E /\alpha_{QED}$,
$P_{TT}\rightarrow 0$, and
$P_{LT}\rightarrow - 2 M \beta_M /\alpha_{QED}$.

Equation~\ref{eq:VCSsig} illustrates that an unpolarized VCS experiment
(including no variation in photon polarization)
can measure only two linear
combinations of the polarizabilities. 
The spin polarizabilities $\gamma_{E2}$ and
$\gamma_{M2}$ enter the unpolarized (NB) VCS amplitude to lowest order
even though they only enter the spin dependent 
RCS amplitude to order
$\omega^3$ (Equation \ref{eq:RCSamp}). 
First, $\gamma_{E2}$ and $\gamma_{M2}$ enter the unpolarized VCS
cross section because the virtual photons are linearly polarized, and
this induces a polarization of the proton 
in the  interfering  BH+Born term.  
Second, in the ${\mathcal H}^3$ terms
of Equation~\ref{eq:SpinPol}, if the space derivatives are assigned to the initial
virtual photon,  these terms enter the VCS amplitude to the same order in
$q'$ as do the ordinary polarizabilities.

\begin{figure}[h!]
\epsfxsize=\textwidth
\centerline{\epsfbox{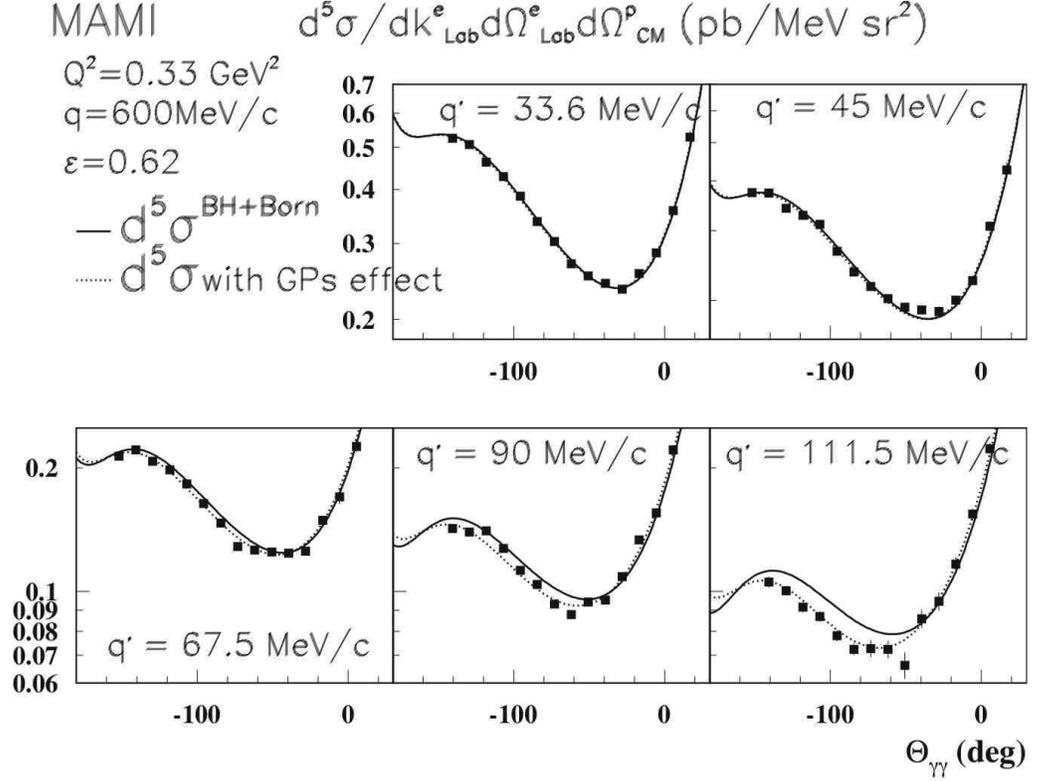}}
\caption{MAMI $ep\rightarrow ep\gamma$ data 
 at fixed $|{\bf q}|=0.6$ GeV and five values of 
$q'$ (both variables in the
photon-proton center-of-mass frame)\cite{Roche:2000ng}.  
The differential cross sections
are plotted as a function of the polar center-of-mass angle $\theta$
between the virtual photon direction 
${\bf q}={\bf k}^e-{\bf k}^{e\prime}$ and the outgoing
photon ${\bf q}^{\prime}$. Values of $\theta<0$ refer to 
kinematics with the  azimuth of 
${\bf q}^{\prime}$ around ${\bf q}$ equal to  $180^\circ$.  
The solid curves are the
BH+Born calculations.  The dashed curves include the low
energy expansion of Equation~\ref{eq:VCSsig}, with the two
structure functions $P_{LL}-P_{TT}/\epsilon$ and
$P_{LT}$ fitted to the data.}
\label{fig:VCSMAMI}
\end{figure}

Roche \etal{}~\cite{Roche:2000ng}{} 
measured the VCS cross
section on the proton below threshold at MAMI
 at fixed ${\rm q}_{CM} = 600$ MeV/c ($\tilde{Q}^2=0.33$ GeV$^2$) and
$\epsilon = 0.62$, where ${\rm q}_{CM}$ is the VCS virtual photon 3-momentum
in the proton-photon center-of-mass frame.
   The cross sections were extracted including 
radiative corrections calculated specifically for the full
VCS process \cite{Vanderhaeghen:2000ws}.
Figure~\ref{fig:VCSMAMI}
displays the angular distributions of the $ep\rightarrow e p\gamma$
cross section for five values of $q'$.  
The rapid
rise in the cross sections for $\theta>0$ is the tail
of the first of the two BH peaks, when the radiated
photon is parallel to the incident (first peak) or scattered electron
direction. The broad rise in the cross section for 
$\theta\le -90^\circ$ has a strong contribution from the
Born term, which is approximately a boosted Larmor 
dipole-radiation pattern from the proton.
The cross section at low $q'$ is consistent with the BH+Born
cross section within the 2.5 \% experimental uncertainty,
including the uncertainty in the elastic proton form factors.
The deviation of the data from the BH+Born cross section grows
linearly with $q'$, as expected for the polarizabilities.

\begin{figure}[h!]
\epsfxsize=13cm
\vskip -7cm
\centerline{\epsfbox{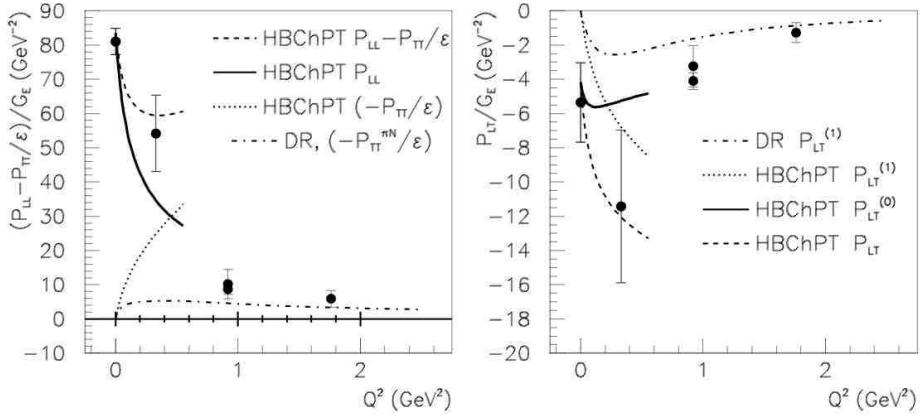}}
\caption{Generalized polarizability structure functions
extracted from
$\gamma p \rightarrow \gamma p$ \cite{OlmosdeLeon:zn}, and 
$ep\rightarrow  ep\gamma$ data of MAMI \cite{Roche:2000ng}{} and
Jefferson Lab \cite{Laveissiere:2003sv}.    
The Jefferson Lab points are the 
low energy expansion and dispersion-relation (DR) set Ib points at $Q^2=1$
and the DR set II point at $Q^2=2$. 
The plots are
$P_{LL}-P_{TT}/\epsilon$ (left) and $P_{LT}$ (right).
The dashed, solid, and dotted curves represent calculations in 
 ${\mathcal O}(p^3)$ Heavy Baryon Chiral Perturbation
Theory (HBChPT) \cite{Hemmert:1999pz}.  
Dashed curves, left and right are the total  HBChPT predictions,
with $\epsilon=0.62$ (MAMI value) at left.
The solid curves at left and right are the  
HBChPT contributions of $\alpha_E$ and
$\beta_M$, respectively.  The dotted curves are the  HBChPT contributions
of the spin polarizabilities $P_{TT}/\epsilon$ (left) and 
$P^{(C1,C1)(1)}$ 
(right).  The dot-dashed curves are the DR predictions for the
same spin polarizabilities \cite{Pasquini:2001yy}.  
All data and curves  are divided by
the Brash parameterization of $G_{E}^p(Q^2)$ \cite{brash}.}
\label{fig:PLLPLT}
\end{figure}

\begin{figure}[h!]
\epsfxsize=\textwidth
\centerline{\epsfbox{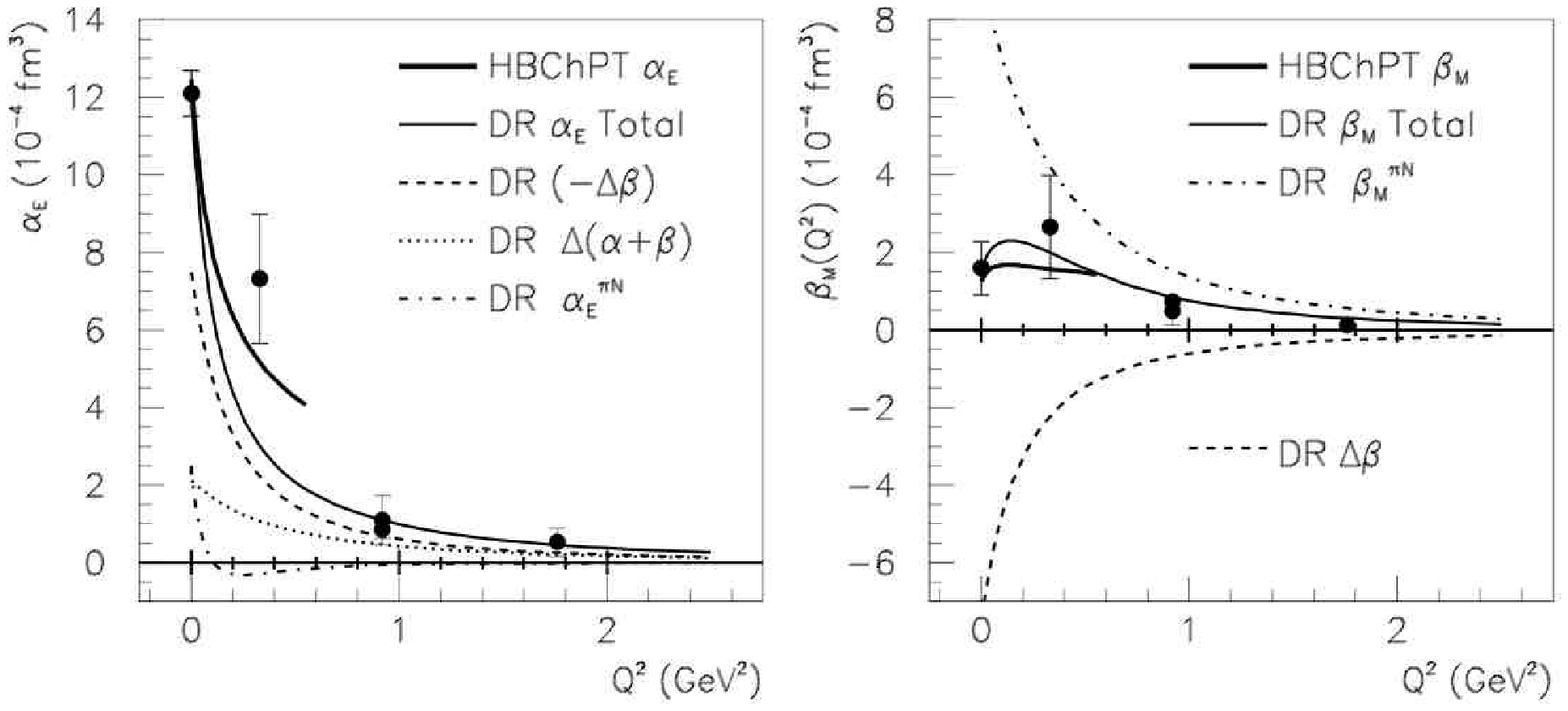}}
\caption{Generalized polarizabilities $\alpha_E$
and $\beta_M$ of the proton.  The (model-dependent) data points
are extracted from Figure~\ref{fig:PLLPLT} by subtracting the DR
predictions of the spin polarizability contributions 
from the data (Eqs.~\ref{eq:PTT},\ref{eq:PLT}).
The heavy solid curves ending at $Q^2$ = 0.5 GeV$^2$ are the
HBChPT predictions for $\alpha_E$ and $\beta_M$ 
(same as Figure~\ref{fig:PLLPLT}) \cite{Hemmert:1999pz}.
See text for discussion of the DR curves.}
\label{fig:alphabeta}
\end{figure}

The generalized polarizabilities have  been calculated
in the Constituent-Quark model 
\cite{Guichon:1995pu,Liu:1996xd,Pasquini:2000ue},
a tree-level  Lagrangian model \cite{Vanderhaeghen:iz},
the Skyrme model \cite{Kim:1997hq}, 
the Linear $\sigma$-model \cite{Metz:fn,Metz:1997fr}
and to ${\mathcal O}(p^3)$ Chiral Perturbation Theory (ChPT)
\cite{Hemmert:1997at,Hemmert:1996gr,Hemmert:1999pz}.
The generalized spin polarizabilities have also been
calculated in ${\mathcal O}(p^4)$ 
HBChPT \cite{Kao:2002cn}.

Figure~\ref{fig:PLLPLT} displays the polarizability structure functions
$P_{LL}-P_{TT}/\epsilon$ and $P_{LT}$ extracted
from the MAMI VCS experiment, along with the RCS results
\cite{OlmosdeLeon:zn} and the Jefferson Lab VCS results
\cite{Laveissiere:2003sv}.  The figure also shows
the  ${\mathcal O}(p^3)$ (one loop)
HBChPT results
\cite{Hemmert:1999pz},
and the DR predictions of the
spin polarizabilities \cite{Pasquini:2001yy}.

Hemmert et al. calculated the generalized polarizabilities 
to ${\mathcal O}(p^3)$ (one loop)
in  HBChPT and to ${\mathcal O}(\epsilon^3)$ in the
small-scale-expansion (SSE) 
 with $M_\Delta-M$ taken
as a third expansion parameter (as well as $m_\pi$ and $p$)
\cite{Hemmert:1999pz}.
The HBChPT calculation yields  analytic expressions for the
generalized polarizabilities, which are plotted in Figure~\ref{fig:PLLPLT}.
The agreement between the calculations and the data at low
$Q^2$ is striking.


The HBChPT calculation of $\beta_M(\tilde{Q}^2)$ has the dramatic feature
of rising at low ${\rm q}$.  As expected in a naive picture,
this results,
from a partial cancellation between the diamagnetism of the pion cloud
and the paramagnetism of the core.  Larger values of ${\rm q}$ probe
shorter distance scales, and are therefore dominated by the paramagnetism.
Eventually the finite size of the proton imposes the decrease of both
the para- and diamagnetic contributions.  

 The strong cancellation between para- and diamagnetism is
also emphasized by both the SSE ${\mathcal O}(\epsilon^3)$ calculation
\cite{Hemmert:1999pz}{} and the tree-level effective Lagrangian
model \cite{Vanderhaeghen:iz}.  The SSE calculation agrees
with the HBChPT calculations for the $Q^2$ variation of the
Generalized Polarizabilities, and for the magnitude of the
spin-polarizabilities.  However, as discussed above in the
RCS section (\ref{sec:RCStheory}), at the photon point:
\begin{eqnarray}
\alpha[SSE\ {\mathcal O}(\epsilon^3)] - 
\alpha[HBChPT\ {\mathcal O}(p^3)] &=& 4.2 \cdot 10^{-4} {\rm fm}^3 
\nonumber\\
\beta[SSE\ {\mathcal O}(\epsilon^3)] - 
\beta[HBChPT\ {\mathcal O}(p^3)] &=& 7.2 \cdot 10^{-4} {\rm fm}^3 
\label{eq:SSE}
\end{eqnarray}
The large value of $\beta_M$ in the SSE calculation comes from the
paramagnetism of the $N\rightarrow \Delta$ transition.
It is expected that this will be canceled by 
${\mathcal O}(\epsilon^4)$ dia-magnetic terms.
Similarly, in the tree-level effective Lagrangian model, there
is a strong cancellation between the $N\rightarrow \Delta$
paramagnetism and the diamagnetic contribution from
higher resonances \cite{Vanderhaeghen:iz}.



Pasquini \etal{} developed a DR formalism for the VCS
amplitude up to the $N\pi\pi$ threshold\cite{Pasquini:2001yy}.  
The imaginary part of the 
VCS amplitude is expressed explicitly by unitarity in terms of the 
$\gamma^* N \rightarrow \pi N \rightarrow \gamma N$ MAID multipoles
\cite{Drechsel:1998hk,Tiator:2003uu}.
The real part of the amplitude is expressed as a dispersive integral
(as a function of $\nu=(s-u)/(4M)$ at fixed $t$ and $Q^2$) 
over the imaginary part, by the Cauchy theorem.  If the dispersive
integrals do not saturate at finite energy, then an
($\nu$-independent) asymptotic piece
is added to the amplitude.  This  represents, equivalently,
either a semicircular
contour in the complex $\nu$-plane or the contribution of channels
beyond $\pi N$.
Of the 12
VCS amplitudes, $F_i(Q^2,\nu,t)$, 
the dispersive integrals converge in principle for
all but $F_1$  and $F_5$, based on Regge phenomenology.  

The asymptotic contribution to $F_5$
is obtained from $t$-channel $\pi^0$-exchange, in accord with the
calculation of the backward spin polarizability $\gamma_\pi$ in RCS.
The asymptotic contribution to $F_1$ is obtained from $t$-channel
$\sigma$-exchange, with a phenomenological $\gamma^*\gamma\sigma$
vertex $\Delta\beta(Q^2)$ that must be fitted to the VCS data.
In this approximation, the non-Born (NB) contribution to the 
amplitude $F_1$ is
expressed as:
\begin{eqnarray}
F_1^{NB}(Q^2, \nu,t) &=& F_1^{\pi N}(Q^2,\nu,t) + F_1^{\rm asy}(Q^2,0,t) 
\nonumber \\ 
&=& F_1^{\pi N}(Q^2,\nu,t) +
\sqrt{2E \over E + M}{\Delta\beta(Q^2) \over \alpha_{QED}}
{1+Q^2/m_\sigma^2 \over 1 - t/m_\sigma^2},
\end{eqnarray}
where $E=\sqrt{M^2+{\bf q}^2}$ is the initial proton energy in the
photon-proton center-of-mass frame in the $q'\rightarrow 0$ limit.
The model dependence arises from the assumption that the 
asymptotic term  $F_1^{\rm asy}$
is independent of $\nu$
(at least below $N\pi\pi$ threshold).
The magnetic polarizability is obtained from 
the $(\nu,t)=(0,-Q^2)$ limit of the $F_1^{NB}$
amplitude:
\begin{eqnarray}
\beta_M(Q^2) &=& F_1^{NB}(Q^2, 0, 0) \alpha_{QED} \sqrt{(E+M)/(2E)}.
\label{eq:F1beta}
\end{eqnarray}
The dispersive integral for $F_2$ converges in principle, but in
practice is poorly saturated by the MAID $\pi N$ multipoles
\cite{Pasquini:2001yy}.

The asymptotic part of $F_2^{NB}$ is the only contribution to  
$[\alpha+\beta](Q^2)$ that is not predicted by the DR calculations.
In the absence of a multipole decomposition of the 
$\gamma N \rightarrow \pi\pi N$ amplitudes, the DR analysis 
approximates the low energy behavior of $F_2$ in terms of the $\pi N$
multipoles plus a $\nu$- and $t$-independent asymptotic term:
\begin{eqnarray}
F_2^{NB}(Q^2,\nu,t) &=& F_2^{\pi N}(Q^2,\nu,t) + F_2^{\rm asy}(Q^2,0,0) 
\nonumber \\
F_2^{NB}(Q^2,\nu,t) &=& F_2^{\pi N}(Q^2,\nu,t)
-\sqrt{2E \over E + M} {1\over 4M^2 (1+\tau)}
{\Delta[\alpha+\beta](Q^2) \over \alpha_{QED} }.
\end{eqnarray}
The polarizability sum is (\cite{Pasquini:2001yy}, Equation~29):
\begin{eqnarray}
\lefteqn{ [\alpha_E+\beta_M](Q^2) = \Delta[\alpha+\beta](Q^2) } \nonumber \\
 & & -4M^2 \alpha_{QED} \sqrt{E+M\over  2 E} \left[
(1+\tilde{\tau})F_2+ 2F_6+F_9-F_{12}\right]^{\pi N}(Q^2,0,-Q^2)
\label{eq:F2Dab}
\end{eqnarray}

In summary, 
the DR formalism of Reference \cite{Pasquini:2001yy}{} 
has a complete prediction of the VCS
amplitude up to $N\pi\pi$ threshold, including all spin polarizabilities,
in terms of just two unknown functions to be extracted from the data:
$\Delta\beta(Q^2)$ and $\Delta[\alpha+\beta](Q^2)$.
At the real photon point, if one combines the 
DR calculations with the
experimental results in Equation~\ref{eq:Olmos} \cite{Pasquini:2001yy}:
\begin{eqnarray}
       \beta_M^{\pi N}(0) &=& +9.1 \cdot 10^{-4} \,{\rm fm}^3 \nonumber \\
           \Delta\beta(0) &=& -7.5 \cdot 10^{-4} \,{\rm fm}^3 \\
\left[ \alpha + \beta \right]^{\pi N}(0) &=& 11.6 \cdot 10^{-4} \,{\rm fm}^3 \nonumber \\
  \Delta[\alpha+\beta ](0) &=& +2.1 \cdot 10^{-4} \,{\rm fm}^3 
\end{eqnarray}
The $\pi N$ DR contribution is seen to be strongly paramagnetic,
which arises naturally from the paramagnetic response of the constituent quarks
which define the resonance spectrum. The asymptotic piece $\Delta\beta$
is required to be strongly diamagnetic, which again is a natural result
given the formal link with the pion cloud of the nucleon (via the
$t$-channel $\sigma=[\pi\pi]_0$ exchange).  Finally, the
asymptotic $F_2$ term $\Delta[\alpha+\beta](0)$
is only $15\%$ of the total Baldin sum rule (Equation~\ref{eq:BB}).

The Jefferson Lab VCS collaboration analyzed $e p \rightarrow e p \gamma$
data below pion
threshold in terms of the low-energy expansion 
(Equation~\ref{eq:VCSsig}) and
data up through the $\Delta$-resonance in terms of the DR
formalism. The results are shown in 
Figures~\ref{fig:PLLPLT} and \ref{fig:alphabeta}.
Although the amplitudes $F_1$ and $F_2$ are the natural degrees
of freedom in the DR formalism, the Jefferson Lab DR analysis follows 
\cite{Pasquini:2001yy} in estimating the asymptotic terms with
two dipole form factors:
\begin{eqnarray}
\Delta\alpha_E &=& {\Delta\alpha(0) 
\over \left[1 + Q^2/\Lambda_\alpha^2\right]^2} \nonumber \\
\Delta\beta_M &=&  {\Delta\beta(0) 
\over \left[1 + Q^2/\Lambda_\beta^2\right]^2}, \label{eq:Lambda}
\end{eqnarray}
with the dipole parameters $\Lambda_\alpha$ and  $\Lambda_\beta$ fitted
to the data at each $Q^2$ point.  The dipole form is not essential to
the analysis, since it is  used only to describe the $Q^2$-dependence
within the acceptance of one spectrometer setting.

In Figure~\ref{fig:alphabeta} the DR predictions of 
the spin polarizabilities are used to extract $\alpha_E(Q^2)$ and 
$\beta_M(Q^2)$ from the data.  These data points are therefore
subject to confirmation of the DR spin polarizabilities.
In these two plots, we show the individual contributions to the
total DR calculations.
In the right-hand plot, the dot-dashed line is
the contribution to $\beta_M$ of the $\pi N$ multipoles.  
The dashed line is the contribution of the phenomenological $\Delta\beta$
term, with $\Lambda_\beta = 0.63$ GeV.  The thin solid line is the complete
DR calculation for $\beta_M(Q^2)$.  In the left-hand  plot, the
dot-dashed line is the contribution to $\alpha_E$ of the  $\pi N$ multipoles
(this has a small dia-electric contribution for $Q^2 >0.1$ GeV$^2$).  
In accord with Equations~\ref{eq:F1beta} and \ref{eq:F2Dab}, we write the
asymptotic contribution as 
$\Delta\alpha_E = \Delta[\alpha+\beta]-\Delta\beta$.  The dashed
curve is the contribution of the $F_1^{\rm asy}$ term $-\Delta\beta$
to $\alpha_E$.  Thus the pion cloud ($t$-channel $\sigma$-meson exchange)
makes a diamagnetic contribution to $\beta_M$ and
a positive contribution to $\alpha_E$.
   The dotted curve is the contribution of the 
$F_2^{\rm asy}$ term, using a dipole form adjusted to the Jefferson Lab data:
\begin{eqnarray}
\Delta[\alpha+\beta](Q^2) &=& 
{\Delta[\alpha+\beta](0) \over [1 + Q^2/\Lambda^2_{\alpha\beta}]^2} \nonumber \\
\Lambda_{\alpha\beta} &=& 0.9 \, {\rm GeV}
\end{eqnarray}

The dipole parameter $\Lambda_\beta = 0.63$ GeV is much smaller than the
standard dipole fit to the nucleon form factors.  This supports the 
interpretation that the pion cloud contribution has a much larger spatial
size in the polarizabilities than in the form factors.
The dipole form for $\Delta[\alpha+\beta]$ is unable to reproduce all of the
data for $\alpha_E$.  In particular, the DR curve falls well below 
the MAMI VCS point.  The $S_{11}(1535)$ and $D_{13}(1520)$ resonances
have  strong $E1$ couplings to the $\eta N$ and $\pi\pi N$ channels,
respectively, which are not
included in the MAID analysis \cite{Drechsel:1998hk,Pasquini:2001yy}.  
Although the $A_{3/2}$
helicity amplitude for the $D_{13}$ resonance falls rapidly with
$Q^2$, the $A_{1/2}$ helicity amplitude rises by a factor of
two from $Q^2 = 0$ to 1.0 GeV$^2$ and the  $A_{1/2}$ helicity amplitude of
the $S_{11}$ resonance falls by less than $20\%$ from
$Q^2 = 0$ to 1.0 GeV$^2$ \cite{Tiator:2003uu,Brasse,Thompson:2000}.
Thus it is very plausible that the $\eta N$ and $\pi\pi N$ channels
contribute strongly to $[\alpha+\beta]$ at short distance, 
as suggested by the data in Figure~\ref{fig:alphabeta}.

\begin{figure}[h!]
\centerline{
\epsfxsize=0.46\textwidth
\epsfbox{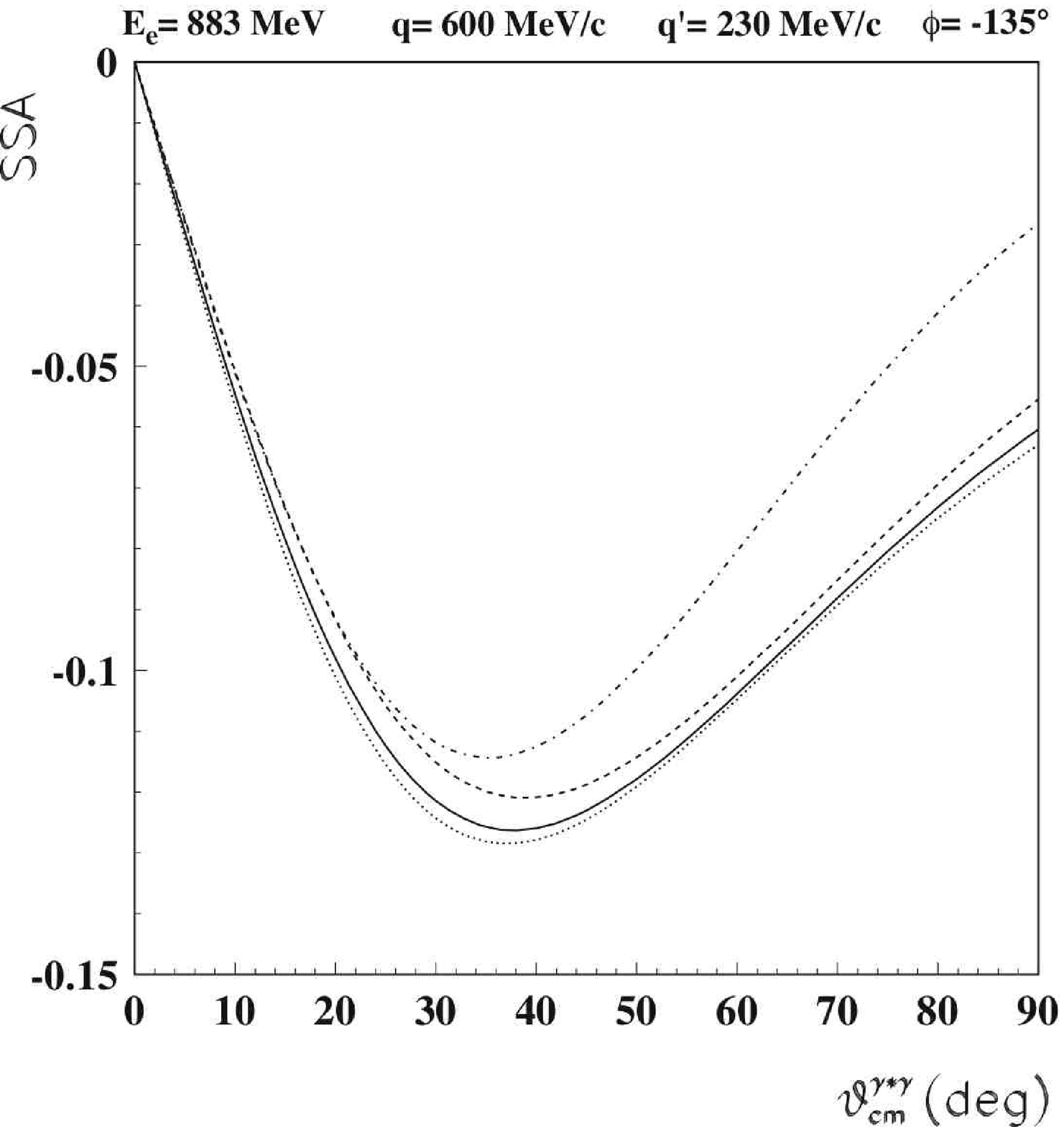}\hfil
\epsfxsize=0.53\textwidth
\epsfbox{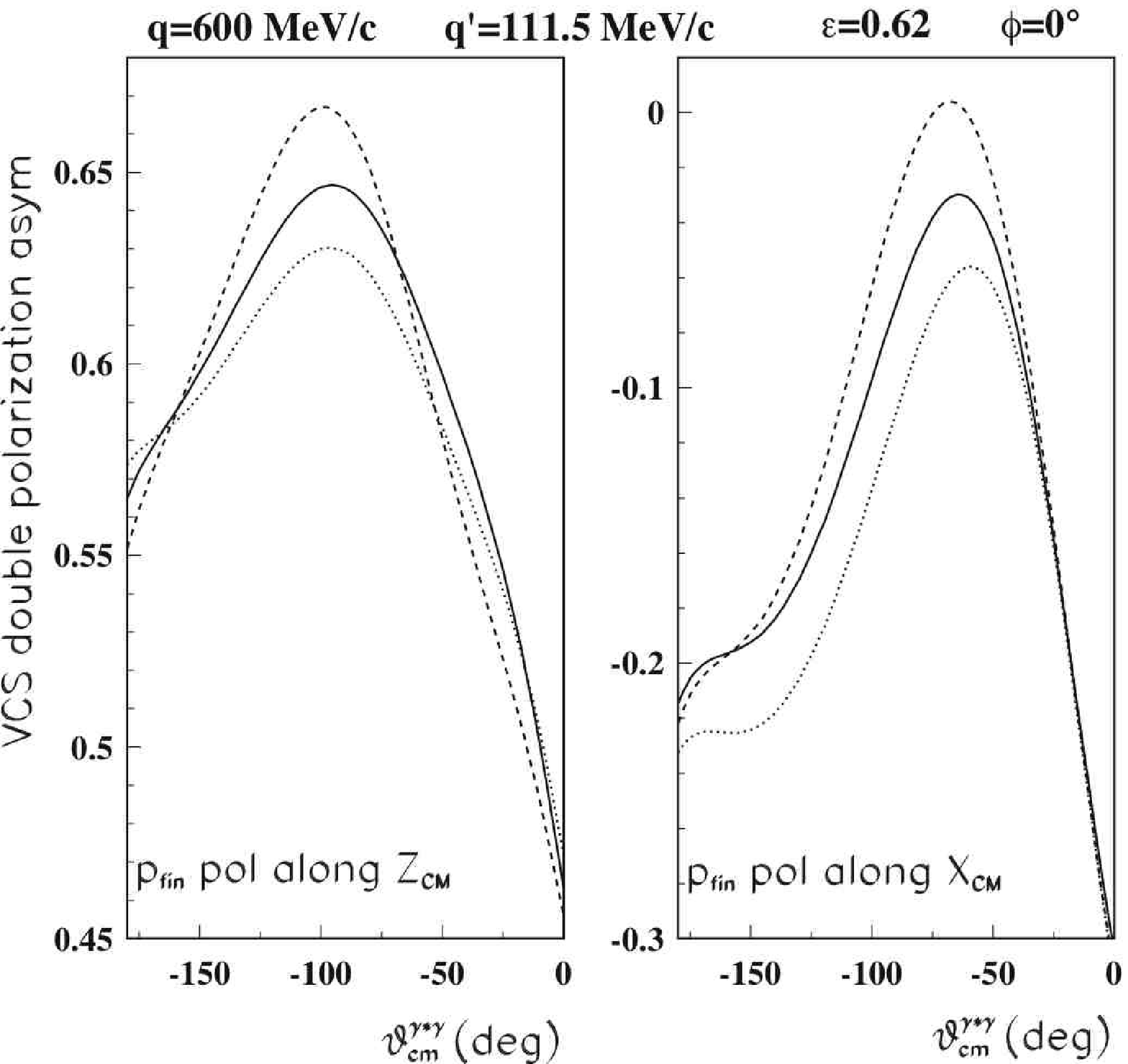}}
\caption{Single and double spin asymmetries in the kinematics of the
MAMI VCS experiment.  Left:  Beam helicity asymmetry in 
H$(\vec{e},e'p)\gamma$.  The four curves are the full 
DR predictions, with the asymptotic terms parameterized by 
$(\Lambda_\alpha,\Lambda_\beta)=(1.0,0.6)$ GeV (solid);
(1.0,0.4) GeV (dashed); (1.0,0.7) GeV (dotted); and 
(1.4,0.4) GeV (dot-dashed).
Right two panels: double-polarization observables in 
H$(\vec{e},e'\vec{p})\gamma$, $Z$-axis parallel to {\bf q}, $X$-axis
in electron scattering plane.  The dotted curves show  the Bethe-Heitler+Born
contribution; the solid curve shows the full DR calculation with
$(\Lambda_\alpha,\Lambda_\beta)$ = (1.0,0.6) GeV (same model as left panel);
the dashed curves represent the HBChPT predictions \cite{Hemmert:1999pz}.}
\label{fig:vcspol}
\end{figure}

The M.I.T.-Bates VCS \cite{ref:Shaw}{} 
experiment used the Out-of-Plane-Spectrometer
(OOPS) system to measure the full azimuthal distribution of recoil
protons around the {\bf q} direction at $Q^2=0.05$ GeV$^2$.  This
gives a strong sensitivity to the polarizabilities.
 This very low $Q^2$ point will
be  valuable in testing chiral dynamics and
assessing the diverse length scales present in
the proton polarization response.

The HBChPT calculations in Figure~\ref{fig:PLLPLT}
suggest that the spin-polarizabilities make
a large contribution to the unpolarized VCS observables, although
the DR predictions indicate otherwise.  This
emphasizes the need for double polarization measurements, that
can directly measure the spin polarizabilities.
These observables were calculated by Vanderhaeghen
\cite{Vanderhaeghen:1997bx}, and are displayed in 
Figure~\ref{fig:vcspol}. 
The Mainz VCS collaboration has completed data taking on both
single and double spin observables in H$(\vec{e},e'p)\gamma$
and H$(\vec{e},e'\vec{p})\gamma$, in the kinematics of 
Figure~\ref{fig:vcspol}.

The spin-polarizability term $P_{TT}$ (Equation~\ref{eq:VCSsig}) can be
isolated with a conventional Rosenbluth separation.
This was discussed in the original Jefferson Lab VCS proposal
\cite{E93050}, and
will be feasible at MAMI with the energy upgrade.
Laveissi\`ere \etal{} show that  higher $Q^2$ measurements
of the polarizabilities are feasible with Jefferson Lab at 6 GeV and above
\cite{Laveissiere}.

\subsection{Inclusive Electron Scattering and Forward Polarizabilities}

The inclusive electron scattering cross section on a nucleon has the
form \cite{Drechsel:2002ar}:
\begin{eqnarray}
 {d\sigma\over dk_{\rm lab}^\prime d\Omega_{\rm lab}} &=& 
 {d\Gamma \over  dk_{\rm lab}^\prime d\Omega_{\rm lab}} 
\sigma(\nu,Q^2) \\
\sigma(\nu,Q^2) &=& \sigma_T +\epsilon \sigma_L
          - h P_x \sqrt{2\epsilon(1-\epsilon)}\, \sigma_{LT}
          - h P_z \sqrt{1-\epsilon^2}\, \sigma_{TT} \\
 {d\Gamma \over  dk_{\rm lab}^\prime d\Omega_{\rm lab}} &=&
{\alpha_{QED} \over 2 \pi^2} \left[{k^{e\prime}\over k^e} \right]^{\rm lab}
{K\over Q^2} {1\over 1-\epsilon}.
\end{eqnarray}
In these expression, $k^e$ and $k^{e\,\prime}$ 
are the incident and scattered electron
energies in the lab frame, and $P_z$ and $P_x$ are the target nucleon
polarizations parallel and perpendicular to {\bf q} (Figure \ref{fig:VCSkin})
in the 
electron scattering plane.
The virtual photon flux $d\Gamma$ is
evaluated in the Hand convention \cite{Hand:bb}: 
$K= (s-M^2)/(2M) = \nu-Q^2/(2M)$, with $\nu=k^e-k^{e\,\prime}$ (in this
sub-section) the electron energy loss.

The partial cross sections are related to the usual DIS structure functions
and the helicity cross sections $\sigma_{1/2}$ and $\sigma_{3/2}$ as follows, 
\begin{eqnarray}
\sigma_T &=& {4\pi^2\alpha_{QED} \over M K} F_1(x_{\rm Bj},Q^2), \nonumber \\
\sigma_L &=& {4\pi^2\alpha_{QED} \over M K} \left[
             \left(1+1/\gamma^2\right)(M/\nu)F_2(x_{\rm Bj},Q^2)
                        -F_1(x_{\rm Bj},Q^2)\right], \nonumber \\
\sigma_{TT} &=& {4\pi^2\alpha_{QED} \over M K} \left[
               g_1(x_{\rm Bj},Q^2) - \gamma^2 g_2(x_{\rm Bj},Q^2)\right]
          = {\sigma_{1/2} - \sigma_{3/2} \over 2},
  \nonumber \\
\sigma_{LT} &=&  {4\pi^2\alpha_{QED} \over M K} \gamma \left[
               g_1(x_{\rm Bj},Q^2) + g_2(x_{\rm Bj},Q^2)\right], 
\label{eq:sigmaDIS}\\
\sigma_{1\pm 1/2} &=& {4\pi^2\alpha_{QED} \over M K}
\left[ F_1(x_{\rm Bj},Q^2) \mp g_1(x_{\rm Bj},Q^2) 
      \pm \gamma^2 g_2(x_{\rm Bj},Q^2) \right], \label{eq:sigHel}
\end{eqnarray}
with  $\gamma^2 = Q^2/\nu^2$, and $x_{\rm Bj} = Q^2/(2M\nu)$ the
Bjorken momentum-fraction variable \cite{Bjorken:1969ja}.
$\sigma_{1\pm 1/2}=\sigma_{3/2}$, $\sigma_{1/2}$ are the
virtual photo-absorption cross sections with total photon+proton
helicity 3/2 and 1/2, respectively.

\begin{figure}[h!]
\begin{center}
\epsfxsize=0.8\textwidth
\epsfbox{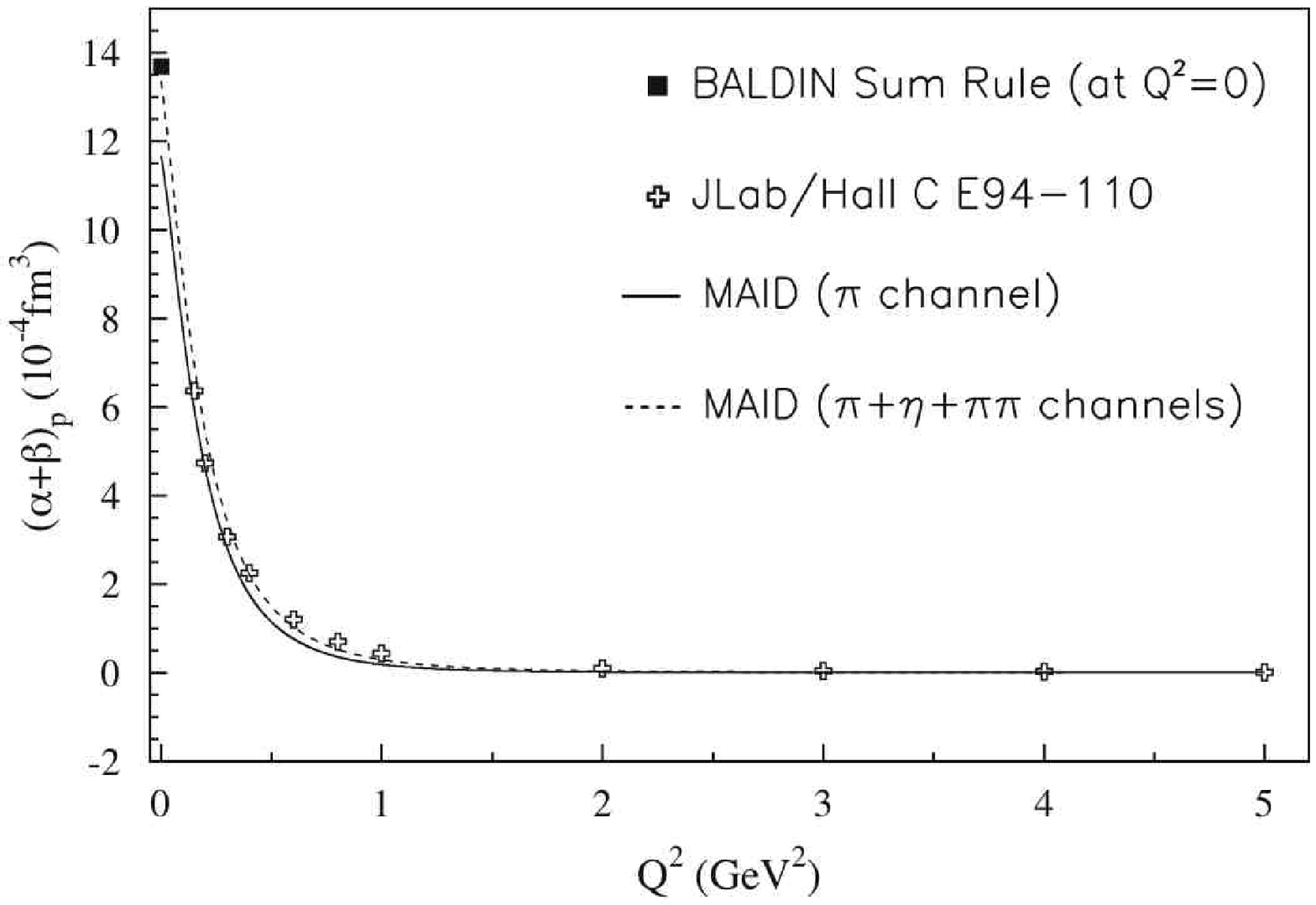}\\
\epsfxsize=0.8\textwidth
\epsfbox{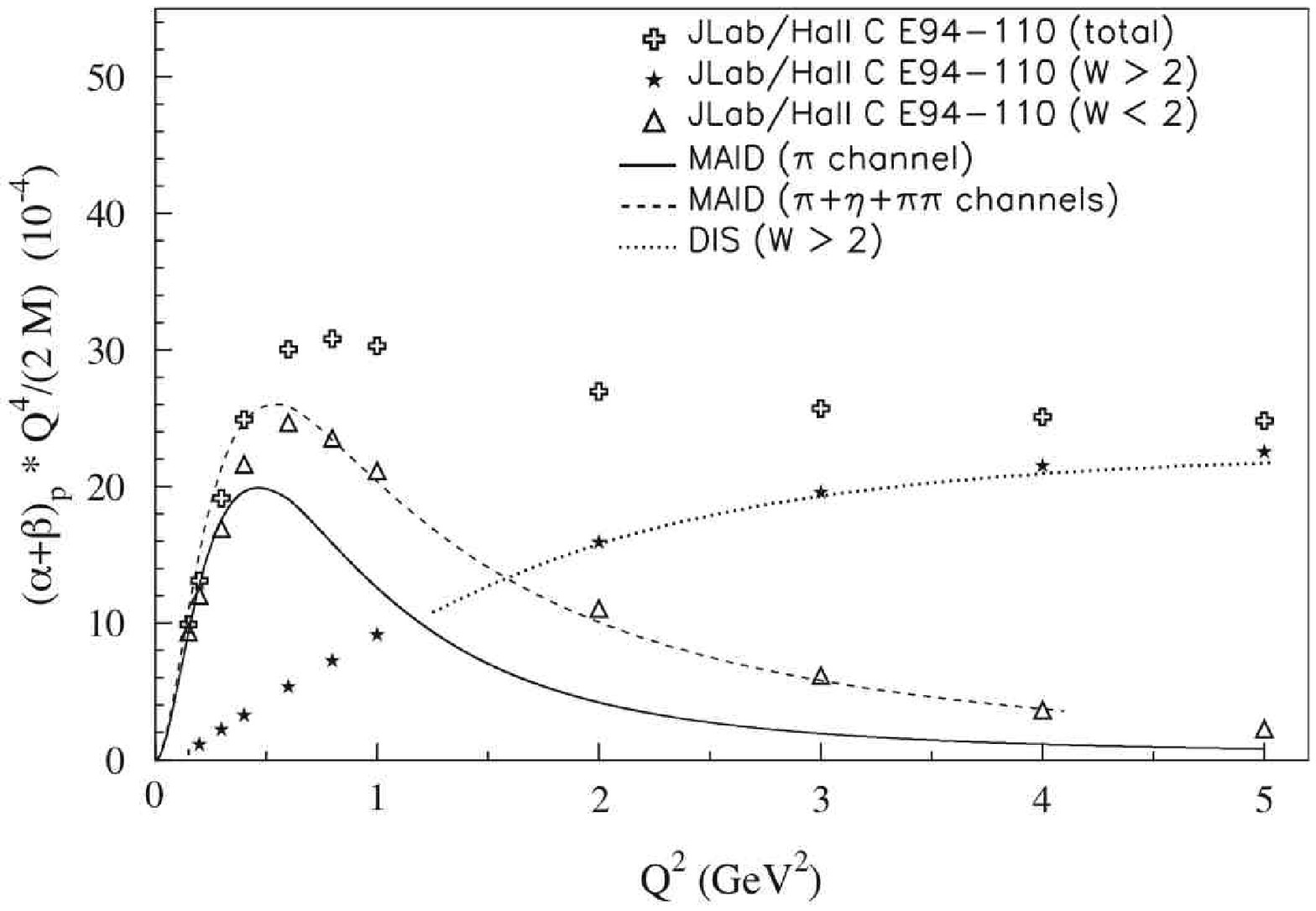}
\caption{$Q^2$-dependence of the generalized forward Baldin sum rule
(Equation~\ref{eq:ab}). The real photon point is from \cite{OlmosdeLeon:zn}, 
the Jefferson Lab data
are from \cite{ref:Liang}.  The curves are taken from  Figure 6 of 
\cite{Drechsel:2002ar}.  The solid curve is from the MAID parameterization of
$\gamma^{(\ast)} N\rightarrow \pi N$ multipoles \cite{Drechsel:1998hk}.
  The dashed 
curve includes the contributions of $\eta N$ and $\pi\pi N$ intermediate
states \cite{Drechsel:2002ar}.  In the lower panel, the dotted curve is 
the estimate from the DIS structure function $F_1$ \cite{Martin:2002dr}, 
for $W >$\ 2 GeV.}
\label{fig:Q2Baldin}
\end{center}
\end{figure}

The inclusive electron scattering cross section is related via the 
optical theorem to the
imaginary part of the forward doubly virtual Compton amplitude 
$\gamma^* N \rightarrow \gamma^* N$.
Starting from a generalization of Equation~\ref{eq:RCSamp} for virtual photons,
Drechsel \etal{} defined dispersion relations between the $(e,e')$
partial cross sections and forward doubly virtual  polarizabilities
\cite{Drechsel:2002ar}
\begin{eqnarray}
\left[\alpha+\beta\right](Q^2,Q^2) &=& {1\over 2\pi^2} \int_{\nu_0}^\infty
        {K(\nu,Q^2)\over \nu} \sigma_T(\nu,Q^2) {d\nu \over \nu^2} \nonumber \\
    &=& {4 M \alpha_{QED} \over Q^4 }
        \int_{0}^{x_0} 2 x F_1(x,Q^2) dx  \label{eq:ab}\\
\alpha_L(Q^2,Q^2)                   &=& {1\over 2\pi^2} \int 
        {K(\nu,Q^2)\over \nu} \sigma_L(\nu,Q^2)  {d\nu\over \nu^2} \nonumber \\
    &=& {16 M^3 \alpha_{QED} \over Q^6} 
         \int dx \left\{
        {Q^2\over 4 M^2}\left[F_2-2 x F_1\right] + x^2 F_2\right\} 
         \label{eq:al} \\
I_A(Q^2) \alpha_{QED}/M^2 &=& {1 \over (2\pi)^2 } \int {K\over \nu} 
\sigma_{TT} {d\nu \over \nu} \label{eq:DHGq2} \\
\gamma_0(Q^2,Q^2) &=& {1\over 2\pi^2} \int_{\nu_0}^\infty
       {K(\nu,Q^2)\over\nu} \sigma_{TT}(\nu,Q^2) {d\nu \over \nu^3} \nonumber\\
     &=& {16 M^2 \alpha_{QED} \over Q^6}
          \int dx x^2 \left[ g_1(x,Q^2)-{x^2\over \tau}
g_2(x,Q^2)\right] \label{eq:Fgam0}\\
\delta_{LT}( Q^2,Q^2) &=& {1\over 2\pi^2} {1\over \sqrt{Q^2}}
         \int_{\nu_0}^\infty
      {K(\nu,Q^2)\over\nu} \sigma_{LT}(\nu,Q^2) {d\nu \over \nu^2} \nonumber \\
    &=& {16 M^2 \alpha_{QED} \over Q^6}
        \int dx x^2  \left[ g_1(x,Q^2) +g_2(x,Q^2)\right] \label{eq:delLT}
\end{eqnarray}
A discussion of the Generalized GDH sum rule $I_A$ (Equation~\ref{eq:DHGq2})
is beyond the scope of this review.  Experimental determination of these
forward polarizabilities requires a full separation of the partial
cross sections in unpolarized and polarized lepton scattering. However,
the polarizability $\alpha+\beta$ dominates the unpolarized cross section,
and at high $Q^2$, $\alpha_L /[\alpha+\beta] \propto 1/Q^2$, illustrating
that $\alpha_L$ is a twist-four matrix element \cite{Drechsel:2002ar}.
The $Q^2\rightarrow 0 $ limits of $\alpha+\beta$ and $\gamma_0$ are 
the usual Baldin
sum rule (Equation~\ref{eq:Baldin}) and the forward spin polarizability
relation
(Equation~\ref{eq:gamma0}).
Using the Wandzura \& Wilczek \cite{Wandzura:qf}{} estimate of $g_2$:
\begin{eqnarray}
\delta_{LT}(Q^2,Q^2) &\rightarrow& \gamma_0(Q^2,Q^2)/3, \ \ \ 
{\rm as} \ Q^2\rightarrow \infty,
\end{eqnarray}

the Jefferson Lab Hall C collaboration has separated the $F_1$ and
$F_2$ structure functions on the proton by the standard Rosenbluth
procedure of measurements at fixed $Q^2$ and variable $\epsilon$
\cite{ref:Liang}. 
Figure~\ref{fig:Q2Baldin} displays the results for the 
forward polarizability $[\alpha+\beta](Q^2,Q^2)$.
The MAID calculation \cite{Drechsel:2002ar,Drechsel:1998hk}, 
with the $\pi N$, $\eta N$, and $\pi\pi N$
intermediate states included
is in good agreement with the data, if the integration
is truncated to $W<2$ GeV.  The estimate from 
DIS structure functions is in good agreement with the data 
if the integrand in both cases is truncated to $W>2$ GeV.
The Jefferson Lab Hall B and C collaborations have also measured the moments
of the $F_2^p(x,Q^2)$ structure function, using previous estimates
of the ratio $R=\sigma_L / \sigma_T$ 
\cite{Osipenko:2003bu,Armstrong:xj}.

\begin{figure}[h!]
\begin{center}
\epsfxsize=0.59\textwidth
\epsfbox{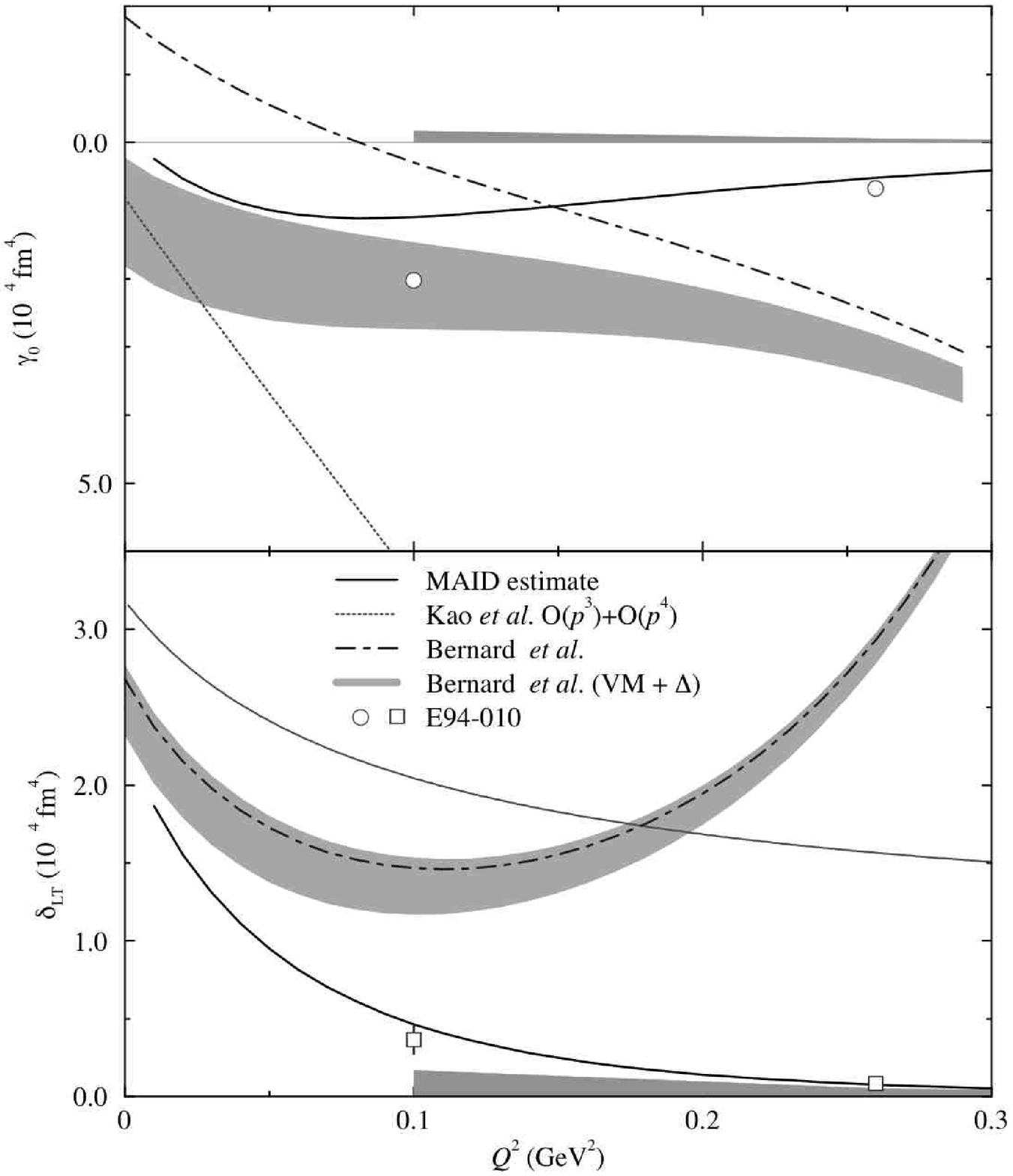}\\
\hbox to \hsize{\hfil
\psfig{file=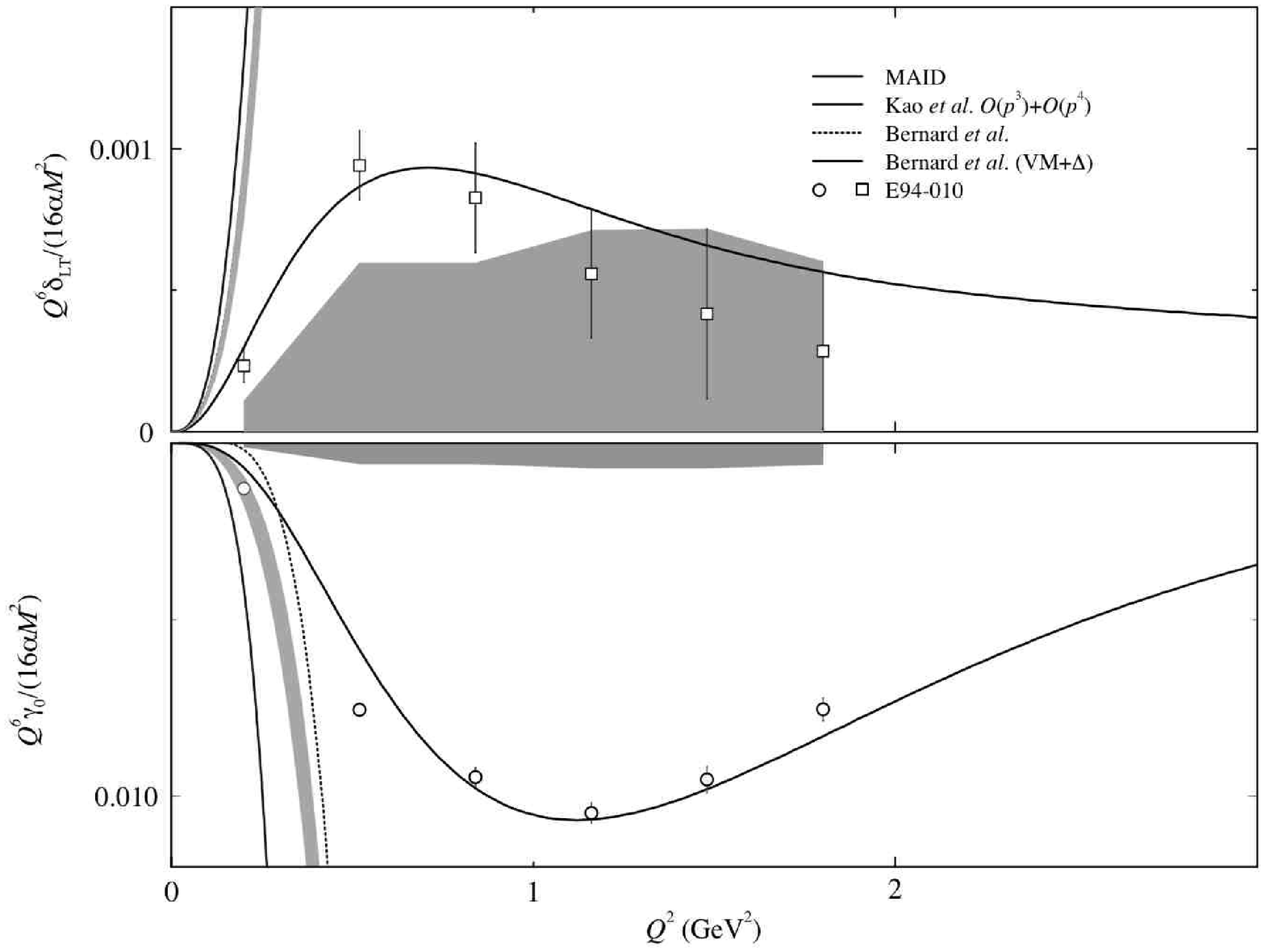,angle=0,width=0.59\textwidth}
\hskip 0.007\textwidth\hfil}
\caption{$Q^2$-dependence of the neutron generalized forward spin
polarizability $\gamma_0^n(Q^2,Q^2)$ (Equation~\ref{eq:Fgam0}) and 
longitudinal-transverse spin polarizability 
$\delta_{LT}^n(Q^2,Q^2)$ (Equation~\ref{eq:delLT}).
The shaded bands on the axes are the systematic error bands of the
Jefferson Lab data \cite{Amarian:2003jy,Amarian:2004}.
The (dark) solid curves extending beyond $Q^2$ =1 GeV$^2$ are 
the MAID parameterization of the $\gamma^* N\rightarrow \pi N$ amplitudes
\cite{Drechsel:2002ar,Drechsel:1998hk}.  The other curves (including the shaded
bands)  are 
ChPT calculations 
(see text for details) \cite{Kao:2002cp,Bernard:2002pw}.
}
\label{fig:Q2deltas}
\end{center}
\end{figure}

The Jefferson Lab Hall A GDH collaboration has  separated
the $g_1^n$ and $g_2^n$ structure functions from measurements of the
$^3\vec{\rm He}(\vec{e},e')$ reaction \cite{Amarian:2003jy}.
Figure~\ref{fig:Q2deltas} displays the forward spin polarizabilities
of the neutron.  The MAID curves under-predict the data for
$\gamma_0$ at low $Q^2$, but are otherwise in good agreement with
the data.  
The curves from  Kao \etal{} are ${\mathcal O}(p^4)$
HBChPT \cite{Kao:2002cp}.  The curves of Bernard \etal{} 
are   ${\mathcal O}(p^4)$ Relativistic Baryon ChPT, with
the shaded bands including the effects of explicit inclusion of the
$\Delta$ and vector mesons \cite{Bernard:2002pw}.  
These explicit resonance effects are large
for $\gamma_0^n$, and bring the ChPT predictions into agreement with the
data at $Q^2=0.1$ GeV$^2$ (but already diverge from the data by
$Q^2=0.25$ GeV$^2$). Although the effect of the resonances is much smaller for
$\delta_{LT}$,  the ChPT predictions disagree sharply with the data in 
Figure~\ref{fig:Q2deltas}.  However, the disagreement is comparable
to the differences between the two ${\mathcal O}(p^4)$ ChPT calculations, 
and the
change from ${\mathcal O}(p^3)$ to ${\mathcal O}(p^4)$
ChPT calculations \cite{Kao:2002cp}.

\section{SUMMARY, OUTLOOK AND CONCLUSIONS}

Recent advances in polarized electron sources, polarized nucleon 
targets and nucleon recoil polarimeters have enabled 
accurate measurements of the spin-dependent elastic electron-nucleon 
cross section. New data on nucleon electro-magnetic form factors with 
unprecedented precision have (and will continue to) become 
available in an ever increasing \Q-domain. The two magnetic form 
factors \GMp\ and \GMn\ closely follow the simple dipole form factor \GD.
\GEpGMp\ drops linearly with \Q\, and \GEn\ appears to drop at the same rate 
as \GEp\ from $\sim 1$ \GeV\ onwards.
The \Q-behavior of \GEp\ has provided a signal of substantial non-zero orbital 
angular momentum in the proton. 
Only scant data are available in the time-like region.
The full EMFF data set forms tight constraints on models of nucleon structure. 
So far, all available theories are at least to 
some extent effective (or parametrizations). Still, only few of these 
adequately describe all four EMFFs. Only lattice gauge theory 
can provide a truely ab initio calculation, but accurate lattice QCD 
results for the EMFFs are still several years away.
A scaling prediction has been developed for the ratio of the Pauli and Dirac 
form factors, which the data appear to follow even at a \Q-value as low as
1 \GeV. Novel procedures allow a visualization
of the nucleon structure as a function of the momentum of the struck quark.
A fully three-dimensional picture of the nucleon will become available
when future exclusive data  have allowed the determination of the
Generalized Parton Distributions.

Measurements of the nucleon polarizabilities have followed a 40-year
odyssey that parallels the history of form-factor measurements.
Powerful new theoretical and experimental techniques  allow high
precision measurements of the polarizablities and their $Q^2$-dependent
generalizations.  These results offer new evidence for the interplay
of constituent quark and pion degrees of freedom at modest distance
scales within the nucleon, and the dynamics of the elementary current
quarks of QCD at large $Q^2$.  This program will continue with both
existing and future experimental facilities.

\section*{Acknowledgments}

It is extremely difficult to write a review on a highly active field; it is impossible to do credit to all important contributions. New papers were submitted while this review was being written. Given the limited space alloted, the best the authors could do was  to briefly summarize a selection of the work that in their opinion had the most impact on the field. We apologize to any author who feels slighted.
This work would have been impossible without the generous help and advice of many individuals. We acknowledge Andrei Afanasev, John Arrington, Hans Hammer, Franco Iachello, Jerry Miller, Antonio Silva, Marc Vanderhaeghen and Barbara Pasquini. The artwork would not have been as good without Rob Feuerbach, Yongguang Liang, Nicole D'Hose, and Geraud Laveissi\`ere.
This work was supported by U.S. Department of Energy contract DE-AC05-84ER40150 Modification No. M175 under which the Southeastern Universities Research Association (SURA) operates the 
Thomas Jefferson National Accelerator Facility. C.E.H.-W. acknowledges support from the U.S. Department of Energy Grant DE-FG02-96ER40960 and thanks its Institute for Nuclear Theory at the University of Washington for its hospitality and partial support during the completion of this work.  C.E.H.-W. also acknowledges
travel grants from the U.S. National Science Foundation and the French Centre National de la Recherche Scientifique.

\end{document}